\documentclass[prd,twocolumn,amsfonts,amssymb,amsmath,eqsecnum,showpacs,letterpaper,superscriptaddress,nofootinbib,altaffilletter]{revtex4}

\usepackage{hyperref}

\usepackage{color}
\usepackage{graphicx}
\usepackage{latexsym}
\usepackage{float}
\usepackage{amsmath}
\usepackage{amssymb}
\usepackage{multirow}
\usepackage{ifthen}

\newcommand{\hrss}{h_{\textrm{rss}}}
\newcommand{\hrssf}{h_{\textrm{rss}}^{50\%}}
\newcommand{\hrssn}{h_{\textrm{rss}}^{90\%}}
\newcommand{\makevisible}[1]{\textcolor{red}{#1}}
\newcommand{\switch}[1]{%
  \ifthenelse{\equal{#1}{0}}{\renewcommand{\makevisible}[1]{}}{}}
\switch{1} 
\def\version$#1,v #2 #3${#2}

\renewcommand{\today}{\number\day\space\ifcase\month\or
  January\or February\or March\or April\or May\or June\or
  July\or August\or September\or October\or November\or December\fi
  \space\number\year}

\def\be{\begin{equation}}
\def\ee{\end{equation}}
\def\bi{\begin{itemize}} 
\def\ei{\end{itemize}}
\def\ben{\begin{enumerate}}
\def\een{\end{enumerate}}

\def\hrss{h_\mathrm{rss}}

\def\ligodoc{{LIGO-P11}{00118}-{v16}} 

\begin{document}



\title{All-sky search for gravitational-wave bursts in the second joint LIGO-Virgo run}

\begin{abstract}
\vspace*{0.2in}
We present results from a search for gravitational-wave bursts
in the data collected by the LIGO and Virgo detectors between
July 7, 2009 and October 20, 2010: data are analyzed when at least two of the three LIGO-Virgo detectors
are in coincident operation, with a total observation time of 207 days.
The analysis searches for transients of duration
$\lesssim 1$ s over the frequency band 64--5000 Hz, without other assumptions on the signal waveform, polarization, direction or occurrence time.
All  identified events are consistent with the expected accidental background. We set frequentist upper limits on the rate of gravitational-wave bursts
 by combining this search with the previous LIGO-Virgo search on the data collected between November 2005 and 
October 2007. The upper limit on the rate of strong gravitational-wave
bursts at the Earth 
is 1.3 events per year at 90\% confidence. We also present upper limits
on source rate density per year and $\mathrm{Mpc^{3}}$ for sample populations of standard-candle sources. 
As in the previous joint run, typical sensitivities of the search in terms of 
the root-sum-squared strain amplitude
for these waveforms lie in the range
$\sim5\times10^{-22}$ Hz$^{-1/2}$ to $\sim1\times 10^{-20}$ Hz$^{-1/2}$. 
The combination of the two joint runs entails the most sensitive all-sky search for generic gravitational-wave bursts and synthesizes the results achieved by the initial generation of interferometric detectors. 
\end{abstract}


\affiliation{LIGO - California Institute of Technology, Pasadena, CA  91125, USA}
\affiliation{California State University Fullerton, Fullerton CA 92831 USA}
\affiliation{SUPA, University of Glasgow, Glasgow, G12 8QQ, United Kingdom}
\affiliation{Laboratoire d'Annecy-le-Vieux de Physique des Particules (LAPP), Universit\'e de Savoie, CNRS/IN2P3, F-74941 Annecy-Le-Vieux, France}
\affiliation{INFN, Sezione di Napoli $^a$; Universit\`a di Napoli 'Federico II'$^b$ Complesso Universitario di Monte S.Angelo, I-80126 Napoli; Universit\`a di Salerno, Fisciano, I-84084 Salerno$^c$, Italy}
\affiliation{LIGO - Livingston Observatory, Livingston, LA  70754, USA}
\affiliation{Albert-Einstein-Institut, Max-Planck-Institut f\"ur Gravitationsphysik, D-30167 Hannover, Germany}
\affiliation{Leibniz Universit\"at Hannover, D-30167 Hannover, Germany}
\affiliation{Nikhef, Science Park, Amsterdam, the Netherlands$^a$; VU University Amsterdam, De Boelelaan 1081, 1081 HV Amsterdam, the Netherlands$^b$}
\affiliation{National Astronomical Observatory of Japan, Tokyo  181-8588, Japan}
\affiliation{University of Wisconsin--Milwaukee, Milwaukee, WI  53201, USA}
\affiliation{University of Florida, Gainesville, FL  32611, USA}
\affiliation{University of Birmingham, Birmingham, B15 2TT, United Kingdom}
\affiliation{INFN, Sezione di Roma$^a$; Universit\`a 'La Sapienza'$^b$, I-00185 Roma, Italy}
\affiliation{LIGO - Hanford Observatory, Richland, WA  99352, USA}
\affiliation{Albert-Einstein-Institut, Max-Planck-Institut f\"ur Gravitationsphysik, D-14476 Golm, Germany}
\affiliation{Montana State University, Bozeman, MT 59717, USA}
\affiliation{European Gravitational Observatory (EGO), I-56021 Cascina (PI), Italy}
\affiliation{Syracuse University, Syracuse, NY  13244, USA}
\affiliation{LIGO - Massachusetts Institute of Technology, Cambridge, MA 02139, USA}
\affiliation{APC, AstroParticule et Cosmologie, Universit\'e Paris Diderot, CNRS/IN2P3, CEA/Irfu, Observatoire de Paris, Sorbonne Paris Cit\'e, 10, rue Alice Domon et L\'eonie Duquet, 75205 Paris Cedex 13, France}
\affiliation{Columbia University, New York, NY  10027, USA}
\affiliation{INFN, Sezione di Pisa$^a$; Universit\`a di Pisa$^b$; I-56127 Pisa; Universit\`a di Siena, I-53100 Siena$^c$, Italy}
\affiliation{Stanford University, Stanford, CA  94305, USA}
\affiliation{IM-PAN 00-956 Warsaw$^a$; Astronomical Observatory Warsaw University 00-478 Warsaw$^b$; CAMK-PAN 00-716 Warsaw$^c$; Bia{\l}ystok University 15-424 Bia{\l}ystok$^d$; NCBJ 05-400 \'Swierk-Otwock$^e$; Institute of Astronomy 65-265 Zielona G\'ora$^f$,  Poland}
\affiliation{The University of Texas at Brownsville and Texas Southmost College, Brownsville, TX  78520, USA}
\affiliation{San Jose State University, San Jose, CA 95192, USA}
\affiliation{Moscow State University, Moscow, 119992, Russia}
\affiliation{LAL, Universit\'e Paris-Sud, IN2P3/CNRS, F-91898 Orsay$^a$; ESPCI, CNRS,  F-75005 Paris$^b$, France}
\affiliation{NASA/Goddard Space Flight Center, Greenbelt, MD  20771, USA}
\affiliation{University of Western Australia, Crawley, WA 6009, Australia}
\affiliation{The Pennsylvania State University, University Park, PA  16802, USA}
\affiliation{Universit\'e Nice-Sophia-Antipolis, CNRS, Observatoire de la C\^ote d'Azur, F-06304 Nice$^a$; Institut de Physique de Rennes, CNRS, Universit\'e de Rennes 1, 35042 Rennes$^b$, France}
\affiliation{Laboratoire des Mat\'eriaux Avanc\'es (LMA), IN2P3/CNRS, F-69622 Villeurbanne, Lyon, France}
\affiliation{Washington State University, Pullman, WA 99164, USA}
\affiliation{INFN, Sezione di Perugia$^a$; Universit\`a di Perugia$^b$, I-06123 Perugia,Italy}
\affiliation{INFN, Sezione di Firenze, I-50019 Sesto Fiorentino$^a$; Universit\`a degli Studi di Urbino 'Carlo Bo', I-61029 Urbino$^b$, Italy}
\affiliation{University of Oregon, Eugene, OR  97403, USA}
\affiliation{Laboratoire Kastler Brossel, ENS, CNRS, UPMC, Universit\'e Pierre et Marie Curie, 4 Place Jussieu, F-75005 Paris, France}
\affiliation{University of Maryland, College Park, MD 20742 USA}
\affiliation{Universitat de les Illes Balears, E-07122 Palma de Mallorca, Spain}
\affiliation{University of Massachusetts - Amherst, Amherst, MA 01003, USA}
\affiliation{Canadian Institute for Theoretical Astrophysics, University of Toronto, Toronto, Ontario, M5S 3H8, Canada}
\affiliation{Tsinghua University, Beijing 100084 China}
\affiliation{University of Michigan, Ann Arbor, MI  48109, USA}
\affiliation{Louisiana State University, Baton Rouge, LA  70803, USA}
\affiliation{The University of Mississippi, University, MS 38677, USA}
\affiliation{Charles Sturt University, Wagga Wagga, NSW 2678, Australia}
\affiliation{Caltech-CaRT, Pasadena, CA  91125, USA}
\affiliation{INFN, Sezione di Genova,  I-16146  Genova, Italy}
\affiliation{Pusan National University, Busan 609-735, Korea}
\affiliation{Australian National University, Canberra, ACT 0200, Australia}
\affiliation{Carleton College, Northfield, MN  55057, USA}
\affiliation{The University of Melbourne, Parkville, VIC 3010, Australia}
\affiliation{Cardiff University, Cardiff, CF24 3AA, United Kingdom}
\affiliation{INFN, Sezione di Roma Tor Vergata$^a$; Universit\`a di Roma Tor Vergata, I-00133 Roma$^b$; Universit\`a dell'Aquila, I-67100 L'Aquila$^c$, Italy}
\affiliation{University of Salerno, I-84084 Fisciano (Salerno), Italy and INFN (Sezione di Napoli), Italy}
\affiliation{The University of Sheffield, Sheffield S10 2TN, United Kingdom}
\affiliation{WIGNER RCP, RMKI, H-1121 Budapest, Konkoly Thege Mikl\'os \'ut 29-33, Hungary}
\affiliation{INFN, Gruppo Collegato di Trento$^a$ and Universit\`a di Trento$^b$,  I-38050 Povo, Trento, Italy;   INFN, Sezione di Padova$^c$ and Universit\`a di Padova$^d$, I-35131 Padova, Italy}
\affiliation{Inter-University Centre for Astronomy and Astrophysics, Pune - 411007, India}
\affiliation{California Institute of Technology, Pasadena, CA  91125, USA}
\affiliation{Northwestern University, Evanston, IL  60208, USA}
\affiliation{University of Cambridge, Cambridge, CB2 1TN, United Kingdom}
\affiliation{The University of Texas at Austin, Austin, TX 78712, USA}
\affiliation{Rochester Institute of Technology, Rochester, NY  14623, USA}
\affiliation{E\"otv\"os Lor\'and University, Budapest, 1117 Hungary}
\affiliation{University of Szeged, 6720 Szeged, D\'om t\'er 9, Hungary}
\affiliation{Rutherford Appleton Laboratory, HSIC, Chilton, Didcot, Oxon OX11 0QX United Kingdom}
\affiliation{Embry-Riddle Aeronautical University, Prescott, AZ   86301 USA}
\affiliation{National Institute for Mathematical Sciences, Daejeon 305-390, Korea}
\affiliation{Perimeter Institute for Theoretical Physics, Ontario, N2L 2Y5, Canada}
\affiliation{University of New Hampshire, Durham, NH 03824, USA}
\affiliation{University of Adelaide, Adelaide, SA 5005, Australia}
\affiliation{University of Southampton, Southampton, SO17 1BJ, United Kingdom}
\affiliation{University of Minnesota, Minneapolis, MN 55455, USA}
\affiliation{Korea Institute of Science and Technology Information, Daejeon 305-806, Korea}
\affiliation{Hobart and William Smith Colleges, Geneva, NY  14456, USA}
\affiliation{Institute of Applied Physics, Nizhny Novgorod, 603950, Russia}
\affiliation{Lund Observatory, Box 43, SE-221 00, Lund, Sweden}
\affiliation{Hanyang University, Seoul 133-791, Korea}
\affiliation{Seoul National University, Seoul 151-742, Korea}
\affiliation{University of Strathclyde, Glasgow, G1 1XQ, United Kingdom}
\affiliation{Southern University and A\&M College, Baton Rouge, LA  70813, USA}
\affiliation{University of Rochester, Rochester, NY  14627, USA}
\affiliation{University of Sannio at Benevento, I-82100 Benevento, Italy and INFN (Sezione di Napoli), Italy}
\affiliation{Louisiana Tech University, Ruston, LA  71272, USA}
\affiliation{McNeese State University, Lake Charles, LA 70609 USA}
\affiliation{Andrews University, Berrien Springs, MI 49104 USA}
\affiliation{Trinity University, San Antonio, TX  78212, USA}
\affiliation{Southeastern Louisiana University, Hammond, LA  70402, USA}
\author{J.~Abadie$^{\text{1},\ast}$}\noaffiliation\author{B.~P.~Abbott$^{\text{1},\ast}$}\noaffiliation\author{R.~Abbott$^{\text{1},\ast}$}\noaffiliation\author{T.~D.~Abbott$^{\text{2},\ast}$}\noaffiliation\author{M.~Abernathy$^{\text{3},\ast}$}\noaffiliation\author{T.~Accadia$^{\text{4},\dagger}$}\noaffiliation\author{F.~Acernese$^{\text{5a,5c},\dagger}$}\noaffiliation\author{C.~Adams$^{\text{6},\ast}$}\noaffiliation\author{R.~Adhikari$^{\text{1},\ast}$}\noaffiliation\author{C.~Affeldt$^{\text{7,8},\ast}$}\noaffiliation
% ...

\collaboration{$^\ast$The LIGO Scientific Collaboration and $^\dagger$The Virgo Collaboration}
\noaffiliation


\maketitle

\section{Introduction}\label{sec:introduction}

Astrophysical sources of transient gravitational waves (duration of 
$\lesssim$ 1 s)~\cite{Cutler:2001} include merging compact binary
systems consisting of black holes and/or neutron stars~\cite{Pretorius:2007nq, Etienne:2007},
core-collapse supernovae~\cite{ott2008}, neutron star 
collapse to black holes~\cite{SNwave}, star-quakes associated with magnetar
flares~\cite{SGR} or pulsar glitches~\cite{glitches}, cosmic string
cusps~\cite{cusp}, and other violent events in the Universe. 
Since many classes of gravitational-wave (GW) bursts cannot be 
modeled well -- if at all -- a search for those sources must be 
sensitive to the widest possible variety of waveforms.

This paper reports on a search for GW bursts occurring during the second joint observation
run of the LIGO~\cite{Abbott:2007kv} and Virgo~\cite{Acernese2006} detectors,
which took place in 2009--2010. 
This search makes 
no prior assumptions on source sky location, signal arrival time, 
or the waveform itself. Event rate upper limits from long-term searches 
of this category have been derived with networks of resonant bar detectors 
with spectral sensitivity limited to around 900 Hz in 1997--2000~\cite{IGEC:2000, IGEC:2003} 
and in 2005--2007~\cite{IGEC:2007, IGEC:2010}. Networks of interferometric detectors  
set more stringent upper limits for GW bursts on a wider bandwidth using the LIGO detectors 
in 2005--2006~\cite{LIGOS4burst,Abbott:2009zi, abbott-2009b}
and during the first joint observation of LIGO and Virgo detectors in 2007~\cite{S5VSR1Burst}.

This second joint LIGO-Virgo search for GW bursts analyzed the frequency band 
spanning 64--5000 Hz.  
We achieved a frequency-dependent sensitivity comparable to or better 
than that of the first joint run, and accumulated $\sim 207$ days of 
observation time interlaced with periods of installing or commissioning 
major hardware upgrades.  Moreover, for the first time a low-latency 
analysis was run with the goal of providing triggers for electromagnetic 
follow-ups of candidates by robotic optical telescopes~\cite{2011arXiv1109.3498T}, radio telescopes,  
and the {\it Swift} satellite~\cite{swift04,2005ApJ...621..558G}. In this paper we focus on the final
results of the GW stand-alone search, which found no evidence for GW bursts.   

This paper is organized as follows.  In Section~\ref{sec:s6detectors} we describe the second joint scientific run: we report on the LIGO and Virgo instrumental upgrades with respect to the first run and on data quality studies.  In Section~\ref{sec:overview} we give a brief overview of the search: the search algorithm, background estimation, the simulations and the calibration uncertainties. The signal models (GW waveforms and source populations) we tested are described in Section~\ref{sec:simulations}. The results of the search are presented in Section~\ref{sec:results}, and astrophysical implications are discussed in Section~\ref{sec:summaries}. The Appendices provide additional details on data characterization and analysis methods.

\section{Second LIGO-Virgo Science Run}\label{sec:s6detectors}

The network of detectors used in this search comprises the two LIGO 4 km 
interferometers, denoted ``H1'' (located in Hanford, WA) and ``L1'' 
(Livingston, LA),
as well as the Virgo 3 km interferometer, denoted ``V1'' (close to Pisa, Italy) 
\footnote{The 2 km detector at the Hanford site (H2) was decommissioned before the second joint LIGO-Virgo run. 
 During previous runs, the latter detector was mainly used to enforce additional event selection criteria by taking advantage of the special relationship for GW signals from the co-located interferometers H1 and H2.
}. 

The LIGO detectors operated from July 7, 2009 to October 20, 2010 in their 
sixth science run (S6).  
The Virgo detector operated from July 7, 2009 to January 8, 2010 
in its second science run (VSR2) and again from August 11 to October 20, 2010 
in its third science run (VSR3).

As in the first joint LIGO-Virgo run~\cite{S5VSR1Burst, S5LowMassLV}, 
the operation of three differently oriented and widely separated 
detectors allows for reasonably complete coverage of the sky for at 
least one gravitational-wave polarization component as well as the 
recovery of some source characteristics such as sky 
location~\cite{cWB-PRCmethod2011, 2011arXiv1109.3498T, Schutz-networks:2011}.

\subsection{Detector Upgrades}

\begin{figure}
\begin{center}
\mbox{
\includegraphics*[width=0.5\textwidth]{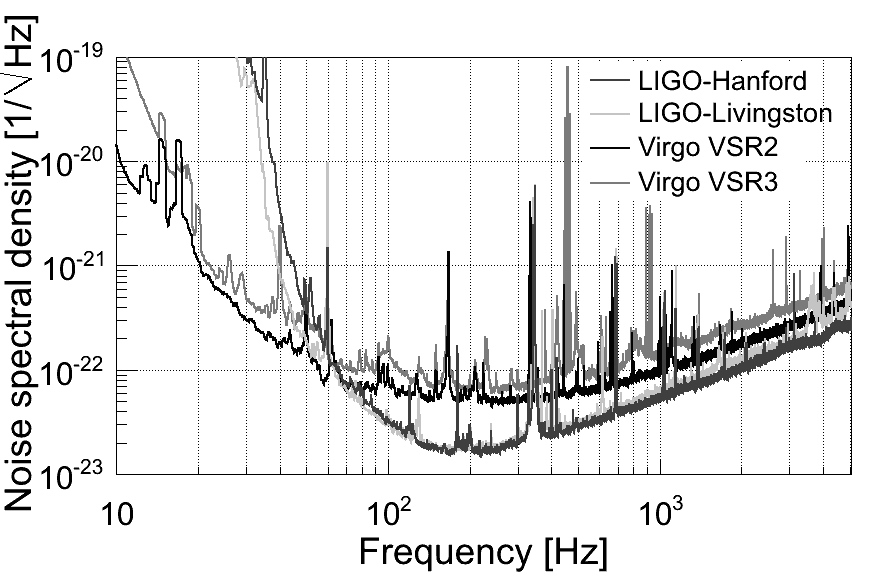}}
\caption{Noise spectra for the three
LSC-Virgo detectors achieved during S6-VSR2/3.}
\label{fig:Shh}
\end{center}
\end{figure}

Before the beginning of the runs, several detector hardware upgrades 
were implemented in order to prototype new subsystems planned for the 
next generation of detectors, referred to as ``advanced 
detectors''~\cite{Harry-advLIGO:2010, Acernese-AdVreport:2009}, expected to start observations in 2015. 
The upgrades of the LIGO detectors for S6 include a higher power 
$35~\mathrm{W}$ laser, the implementation of a DC readout system, a new 
output mode cleaner, and an advanced LIGO seismic isolation table~\cite{Adhikari:2006}. 
The upgrades of the Virgo detectors were achieved in two steps. For VSR2, 
Virgo operated with a more powerful laser and a thermal compensation system. 
Virgo then went offline to undertake a major upgrade of its suspensions, 
mainly in the installation of new test masses consisting of mirrors hung from 
fused silica fibers~\cite{lorenzini2010monolithic}. Virgo resumed 
observations in August 2010 with VSR3. 
Best sensitivities, in terms of noise spectral densities, of the LIGO and Virgo 
detectors achieved during their second joint run (henceforth defined as S6-VSR2/3), as a function of signal 
frequency, are shown in Figure~\ref{fig:Shh}.

\subsection{Data Quality}

To mitigate the consequences of new hardware installations and detector commissioning during this run, significant effort has been made to identify and characterize instrumental
or data acquisition artifacts, periods of degraded
sensitivity, or an excessive rate of transient noise
due to environmental conditions \cite{VirgoDetChar}.
During such times, the data were tagged with Data Quality
Flags (DQFs). Following the same approach used in previous searches~\cite{Abbott:2009zi, abbott-2009b, S5VSR1Burst}, these DQFs are divided into three categories
depending on their impact on the search and on the understanding of the behavior of the detector.
A further description of DQF categories is presented in Appendix \ref{sec:dataquality}.

After DQFs have been applied, the total analyzable time for the 
S6-VSR2/3 run is 242.8 days for H1, 220.2 days for L1, and 187.8 days for V1. 

\section{Search overview}\label{sec:overview}

In this analysis, we considered all four available detector network configurations:
the three detector network, H1L1V1, and the three combinations of detector pairs, H1L1, H1V1 and L1V1. 
We decided \textit{a priori} to search for GW bursts in the entire available time of three-fold observation and 
in the remaining exclusive times of the two-fold networks. 
Table \ref{tab:datatime} reports the total (non-overlapping) coincident observation time for each configuration of detectors
searched for GW signals.  Information about distinct sub-periods of the run may be found in Appendix~\ref{sec:Ltimes}.

Due to the commissioning breaks and installation activities described in Section \ref{sec:s6detectors}, the total observation time is dominated by 2-fold configurations.

\begin{table}[htbp]
\begin{tabular}{|c|c|c|c|c|c|}
\hline
network      & H1L1V1 & H1L1 & L1V1 & H1V1 & total\\
\hline
observation time [days] & 52.2   & 84.5 & 28.9 & 41.0 & 206.6\\
\hline
\end{tabular}

\caption{\label{tab:datatime} Mutually exclusive observation time for each detector
configuration after the application of category 2 DQFs (see Appendix \ref{sec:dataquality} for the definition of data quality flags and their categories). 
}
\end{table}

The useful frequency band is limited to 64--5000 Hz by the
sensitivity of the detectors and by the valid range of data calibration. 
For computational reasons, the event search was performed separately in two suitable bands, 64--2048 Hz and 1600--5000 Hz, 
overlapping to preserve sensitivity to events with spectral power at intermediate frequencies.
The analysis of the events (including the tuning of the search) was performed independently on each configuration of detectors and on three sub-bands, namely 
64--200 Hz, 200--1600 Hz and 1600--5000 Hz, by classifying the 
found events according to their reconstructed central frequency. The motivation for this band splitting is to tune the search within event sets of 
homogeneous glitch behaviour.

\subsection{Search Algorithm}

This search is based on the coherent WaveBurst (cWB) algorithm \cite{Klimenko:2008fu}, which has been used since LIGO's fourth science run in various searches for transient GWs~\cite{Abbott:2008eh, Abbott:2009zi,abbott-2009b, S5VSR1Burst}. 

The cWB analysis is performed in several steps. First, detector data is decomposed into 
a time-frequency representation and then whitened and conditioned to remove 
narrow-band noise features. Events are identified by clustering time-frequency pixels with 
significant energy which is coherent among detectors and characterized using test statistics derived from 
the likelihood (which is also a measure of the signal energy detected in the network and is calculated as 
described in \cite{Abbott:2009zi}). The primary statistics are the network correlation coefficient 
${cc}$, which is a measure of the degree of correlation between the detectors, 
and the coherent network amplitude $\eta$, which is proportional to the 
signal-to-noise ratio and is used to rank events within a homogeneous sub-period. 

Both of these statistics are described in detail in 
\cite{Abbott:2009zi}. The application of the event selection criteria is  thoroughly described in \cite{S5VSR1Burst, Pankow:2009lv}.

Any gravitational-wave candidate event detected by cWB is subject to 
additional data-quality vetoes based on statistical correlations between 
the GW data channel and environmental and instrumental auxiliary channels; 
a significant correlation indicates the event may have been produced by 
local noise. Further details can be found in Appendix \ref{sec:vetoes} .

\subsection{Background Estimation and Search Tuning}

A sample of ``off-source'' (background) events is required to determine the 
selection thresholds to reject noise-induced events contaminating the ``on-source'' 
(foreground) measurement. We estimate the distribution of background events by 
performing the analysis on time-shifted data, typically in $\sim 1$ s steps. 
The shifts minimize the chance of drawing an actual GW into the background sample. To 
accumulate a sufficient sampling, this shifting procedure is performed hundreds or 
thousands of times without repeating the same relative time shifts among detectors.
Background events corresponding to times which are flagged by data quality studies are discarded, just as an event candidate from the foreground would be.

Due to the different characteristics of the background noise for the 
various sub-periods between commissioning breaks and for the different 
frequency bands and networks, 
the thresholds on $cc$ and $\eta$ are tuned separately for each homogeneous sub-period.
Moreover, we consider the action of conditional DQFs (Category 3 DQFs, see 
Appendix \ref{sec:dataquality}) on the event significance, by introducing 
a new ranking scheme which assigns lower significance to events flagged 
by such DQFs. More details on this procedure are reported in Appendix \ref{sec:MIFAR}.

The thresholds reported in Table \ref{Tab:tuning} are selected to require
a false alarm rate (FAR) $\sim$ 1/(8 yr) or lower per frequency
band (to be considered as a “warning threshold”), as described
in Appendix \ref{sec:MIFAR}. 
By considering the union of all searches performed (network 
configurations, sub-periods, and frequency bands) we estimate an overall
false alarm probability (FAP) of $\sim 15\%$. 

The method to measure the statistical significance
of any ``on-source'' candidate GW event which passes
the aforementioned false alarm thresholds was decided {\it a priori}, 
namely performing additional independent time slides to increase 
the statistics of the background estimates.

\subsection{Simulated Signals and Detection Efficiencies}\label{sec:simulations}

\begin{figure*}[!htbp]
 \includegraphics[width=1.01\textwidth]{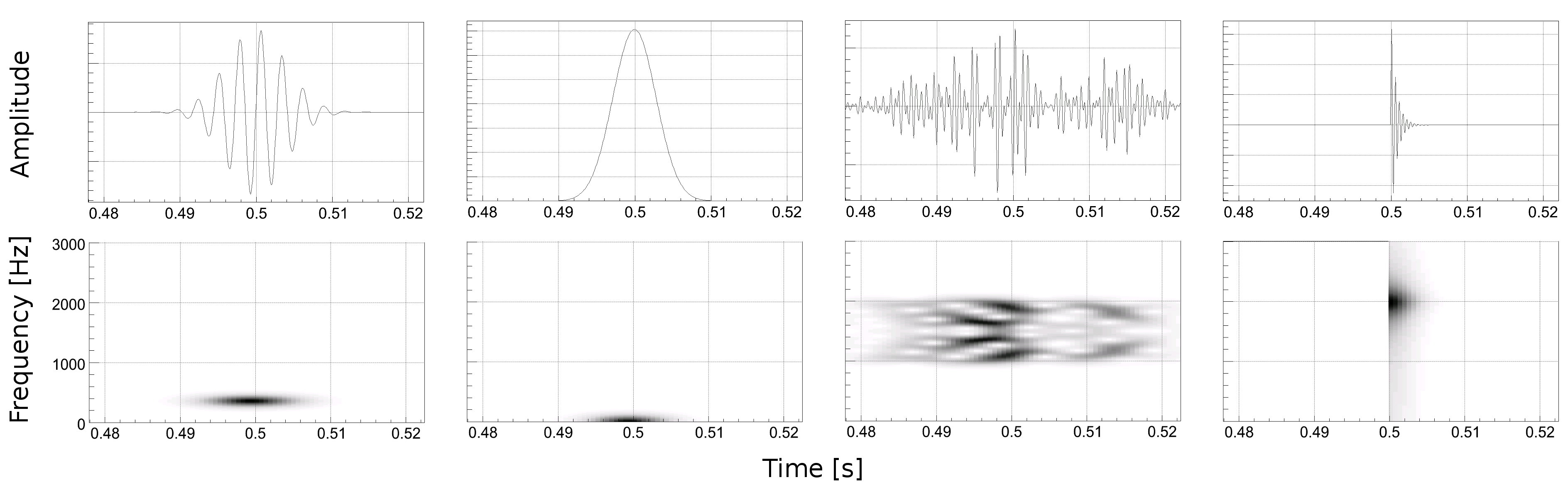}
\caption{Representative waveforms injected into data for simulation studies. The top row is the time domain and the bottom row is a time-frequency domain representation of the waveform. From left to right: a 361 Hz $Q=9$ sine-Gaussian, a $\tau=4.0$~ms Gaussian waveform, a white noise burst with a bandwidth of 1000--2000 Hz and characteristic duration of $\tau=20$~ms and, finally, a ringdown waveform with a frequency of 2000 Hz and $\tau=1$~ms.  
}
\label{fig:waveforms}
\end{figure*}

In order to test the sensitivity of our search to gravitational-wave bursts, we add (``inject'') various ad-hoc software signals, both polarized and un-polarized, to the detector data and measure the detection efficiencies of the search. 
The injected waveforms can be parametrized as: 

\begin{multline}\label{eq:ell}
  \begin{bmatrix}
    h_+ (t)\\ h_\times (t) 
  \end{bmatrix} =
  A \times 
  \begin{bmatrix}
    \frac{1+\alpha^{2}}{2}  \\    \alpha
  \end{bmatrix}
\times 
 \begin{bmatrix}
   H_+(t) \\ H_\times(t) 
  \end{bmatrix},
\end{multline}
where $A$ is the amplitude, $\alpha$ the ellipticity \footnote{For binary sources, the ellipticity is the cosine of the source inclination angle, i.e. the angle between the source rotational axis 
and the line of sight to Earth. } and $H_{+/\times}$ are the waveforms for the two independent polarizations. In this search, we investigated elliptically 
polarized signals (i.e. $\alpha$ uniformly chosen in $[0,1]$), as well as sets of only linearly or circularly polarized waves ($\alpha$ fixed to $0$ or $1$, respectively).  
A variety of GW signal morphologies spanning a wide range of signal durations, frequencies and amplitudes were tested. See Figure \ref{fig:waveforms} for a sample of representative waveforms from various families and Tables \ref{table:SGQ9}, \ref{table:RD}, \ref{table:WNB} for the chosen waveform parameters.   

The injected waveform families include:
\begin{itemize}
\item \textit{Sine-Gaussian}: 
\begin{align}
H_+ (t) &= \exp {(-t^2/\tau^2)} \sin(2\pi f_0 t) \\
H_\times (t) &= \exp {(-t^2/\tau^2)} \cos(2\pi f_0 t)
\end{align}
where $\tau=Q/(\sqrt{2}\pi f_0)$. We consider waveforms of this type 
with central frequencies $f_0$ chosen between 70 to 5000~Hz and quality 
factors $Q=3,9,100$.
Sine-Gaussian waveforms with a few cycles are
qualitatively similar to signals produced by the  mergers of two black holes ~\cite{Pretorius:2007nq}.

\item \textit{Gaussian}:
\begin{align}
H_+(t) &= \exp {(-t^2/\tau^2)} \\
H_\times(t) &= 0
\end{align} 
where the duration parameter $\tau$ is chosen to be one of 0.1, 1.0, 2.5, or 4.0~ms.

\item \textit{Ring-down waveforms}:
\begin{align}
H_+(t) &= \exp(-t/\tau) \sin(2\pi f_0 t) \\
H_\times(t) &= \exp(-t/\tau) \cos(2\pi f_0 t)
\end{align}
We use several central frequencies from 1590~Hz to 3067~Hz, and decay times $\tau = 0.2$~s or $Q=9$. Ring-downs can occur in the end stages of black hole binary mergers. Longer duration ring-downs are also similar to signals 
predicted from the excitation of fundamental modes in neutron stars \cite{Ferrari2004}.

\item \textit{Band-limited white noise signals}:\\
The polarization components are bursts of uncorrelated band-limited white noise, time shaped with a Gaussian profile;   
$H_+$ and $H_\times$ have --- on average --- equal RMS amplitudes and symmetric shape about the central frequency (see Figure \ref{fig:waveforms}).

\item \textit{Neutron star collapse waveforms}: \\
For a comparison with previous searches \cite{abbott-2009b,S5VSR1Burst}, we considered 
numerical simulations by Baiotti et al.~\cite{SNwave}, who modeled neutron
star gravitational collapse to a black hole and the
subsequent ring-down. 
As in previous searches, we chose the models 
D1 (a nearly spherical 1.67 M${}_\odot$
neutron star) and D4 (a 1.86 M${}_\odot$ neutron star that is
maximally deformed at the time of its collapse into a
black hole) to represent the
extremes of the parameter space in mass and spin considered
in the aforementioned work. 
Both waveforms are linearly polarized ($H_\times$ = 0) and their emission is peaked at a few kHz.

\end{itemize}

The simulated signals were injected with many amplitude scale factors to trace out the detection efficiency as a function of signal strength.
The amplitude of the signal is expressed in terms of the root-sum-square strain amplitude ($\hrss$) arriving at the Earth, defined as:

\begin{equation}
\label{eqn:hrss}
\hrss=\sqrt{\int|h_+(t)|^2 + |h_\times(t)|^2dt}
\end{equation}  

The signal amplitude at a detector is modulated by the detector antenna pattern functions, expressed as follows:
\begin{equation}
h_{\text{det}}(t)=F_+(\Theta,\Phi,\psi) h_+(t) + F_\times(\Theta, \Phi,\psi) h_\times(t)
\end{equation}

\noindent where $F_+$ and $F_\times$ are the antenna pattern functions, which depend on the orientation of the wavefront relative to the detector, denoted here in terms of the sky position ($\Theta, \Phi$), and on the
polarization angle $\psi$. The sky positions of simulated signals are distributed 
isotropically and polarization angles are chosen to be uniformly distributed. 
The detection efficiency is defined as the fraction of signals successfully recovered
using the same selection thresholds and DQFs as in the actual search.  
The detection efficiency of the search depends on the network configuration and the selection cuts used in the analysis.

\begin{figure}[!hbtp]
\begin{center}
\includegraphics*[width=0.5\textwidth]{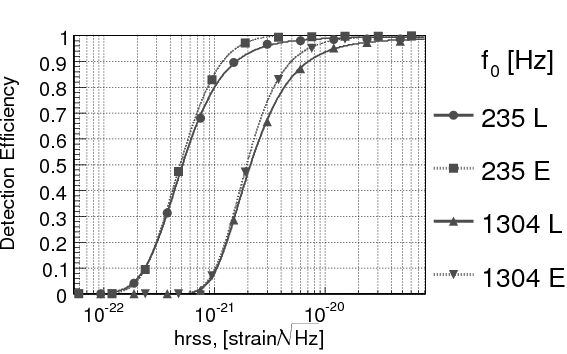}
\includegraphics*[width=0.5\textwidth]{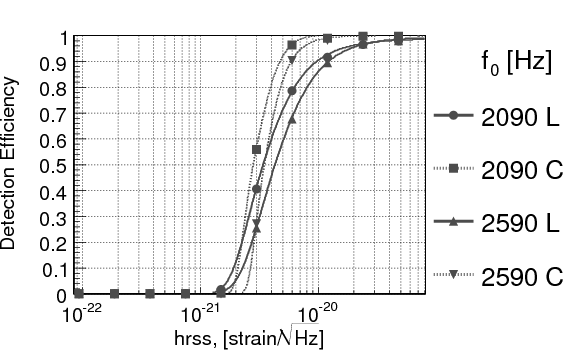}
\includegraphics*[width=0.5\textwidth]{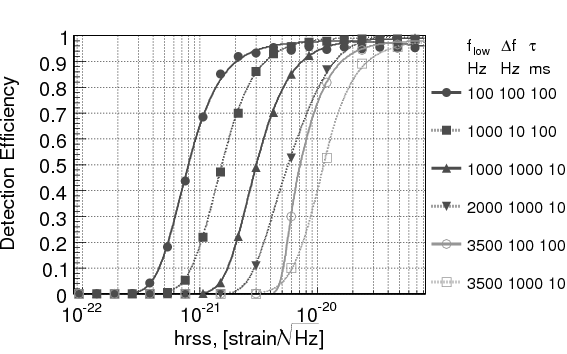}

\caption{
Detection efficiency for selected waveforms as a function of signal amplitude $h_\text{rss}$ for the H1L1V1 network.
Top: Comparison of detection efficiency for linear (L) and elliptical (E) sine-Gaussians with central frequencies of 235 and 1304~Hz.
Middle: Comparison of detection efficiency for linear (L) and circular (C) ring-down signals with frequencies of 2090 and 2590 Hz.
Bottom: Detection efficiency for white noise bursts with frequency spanning between 100 and 4500 Hz.
}
\label{fig:Eff-S6-SineGaussian}
\end{center}
\end{figure}

\begin{figure}[!hbtp]
\begin{center}
\includegraphics*[width=0.5\textwidth]{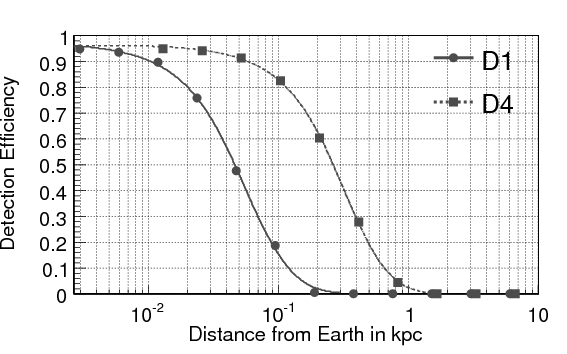}
\caption{Efficiency for the H1L1V1 network as a function
of distance for the D1 and D4 waveforms
predicted by polytropic general-relativistic models of neutron
star collapse.
}
\label{fig:Distance-S6-D1D4}
\end{center}
\end{figure}

Detection efficiencies for the H1L1V1 network for selected waveforms as a function of signal amplitude $h_\text{rss}$  and as a function of distance (for the D1 and D4 waveforms from Baiotti et al.~\cite{SNwave}) are reported in Figures 
\ref{fig:Eff-S6-SineGaussian} and \ref{fig:Distance-S6-D1D4} , respectively.
As in the previous joint run, typical sensitivities for this network in terms of $\hrss$ for the selected waveforms 
 lie in the range
$\sim5\times10^{-22}$ Hz$^{-1/2}$ to $\sim1\times 10^{-20}$ Hz$^{-1/2}$;
typical distances at $50\%$ detection efficiency for neutron star collapse waveforms lie in the range $\sim50$ pc to $\sim200$ pc. 

Two convenient characterizations of the sensitivity, the $\hrss$ at 50\% and 90\% detection efficiency ($\hrssf$ and $\hrssn$ respectively) are obtained from fitting the efficiency curves and are reported in Tables \ref{table:SGQ9}, \ref{table:RD}, and \ref{table:WNB} for the various families. Notice that the 3-fold network, H1L1V1, has a better sensitivity than the weighted average over all networks: 2-fold networks have $\sim 3/4$ of the analyzed live time, but feature a lower sensitivity. 

\begin{table}
  \centering
  \begin{tabular}{cc|c|c|cc|cc}\hline\hline
 &   & \multicolumn{2}{c}{H1L1V1}  & \multicolumn{4}{c}{all networks }\\ 
$f_0$ & $Q$  & Linear   &  Elliptical & \multicolumn{2}{c|}{Linear}   &  \multicolumn{2}{c}{Elliptical}\\ 
 $[$Hz$]$  &   &  $h_{\text{rss}}^{50\%}$ &  $h_{\text{rss}}^{50\%}$ &  $h_{\text{rss}}^{50\%}$ & $h_{\text{rss}}^{90\%}$& $h_{\text{rss}}^{50\%}$ & $h_{\text{rss}}^{90\%}$\\ \hline
70 &  3 & 18.9 & 18.0 & 28.4 & 311.9 & 23.2 & 92.7\\
70 &  9 & 21.5 & 20.4 & 31.6 & 269.4 & 25.8 & 91.7\\
70 &  100 & 24.2 & 21.4 & 34.4 & 484.9 & 27.4 & 131.9\\
100 &  9 & 10.5 & 9.6 & 15.6 & 156.6 & 12.6 & 57.6\\
153 &  9 & 6.7 & 5.8 & 10.3 & 105.4 & 8.0 & 35.2\\
235 &  3 & 5.7 & 5.5 & 8.5 & 45.3 & 7.4 & 24.4\\
235 &  9 & 5.2 & 4.9 & 7.7 & 39.7 & 6.6 & 20.7\\
235 &  100 & 4.6 & 4.4 & 7.2 & 37.6 & 6.0 & 19.0\\
361 &  9 & 8.6 & 8.7 & 12.4 & 67.8 & 11.1 & 32.7\\
554 &  9 & 8.9 & 8.4 & 13.1 & 69.4 & 11.1 & 35.2\\
849 &  3 & 15.1 & 14.4 & 20.8 & 128.7 & 18.4 & 56.6\\
849 &  9 & 14.1 & 13.3 & 19.7 & 116.0 & 17.2 & 52.0\\
849 &  100 & 12.3 & 11.4 & 17.4 & 88.7 & 14.8 & 44.9\\
1053 &  9 & 16.9 & 17.5 & 24.2 & 133.5 & 21.9 & 63.9\\
1304 &  9 & 21.1 & 19.7 & 30.4 & 177.9 & 25.3 & 78.6\\
1615 &  3 & 41.6 & & 54.5 & 349.8\\
1615 &  9 & 35.2 & & 46.3 & 259.9\\
1615 &  100 & 28.3 & & 38.8 & 219.3\\
1797 &  9 & 26.8 & & 35.4 & 206.0\\
2000 &  3 & 41.6 & & 51.8 & 322.9\\
2000 &  9 & 30.8 & & 38.7 & 229.1\\
2000 &  100 & 27.4 & & 36.0 & 181.8\\
2226 &  9 & 36.6 & & 47.2 & 272.1\\
2477 &  3 & 51.6 & & 61.2 & 425.9\\
2477 &  9 & 44.3 & & 55.2 & 307.3\\
2477 &  100 & 34.6 & & 46.3 & 233.5\\
2756 &  9 & 44.2 & & 56.8 & 389.8\\
3067 &  3 & 74.1 & & 81.7 & 600.0\\
3067 &  9 & 64.6 & & 78.0 & 499.6\\
3067 &  100 & 41.1 & & 53.8 & 278.2\\
3413 &  9 & 65.7 & & 80.0 & 510.4\\
3799 &  9 & 81.7 & & 99.3 & 719.9\\

\hline\hline
\end{tabular}
\caption{Values of $h_{\text{rss}}^{50\%}$ and $h_{\text{rss}}^{90\%}$ (for 
50\% and 90\% detection efficiency at the chosen thresholds of 1/(8 yr) per frequency band),
in units of $10^{-22} \, {\rm Hz}^{-1/2}$,
for linear and elliptical sine-Gaussian waveforms with the central frequency $f_0$ and quality factor $Q$. 
The center two columns are the $h_{\text{rss}}^{50\%}$ for linear and elliptical waveforms 
during the total S6 period measured  for the H1L1V1 network.  
The rightmost columns report the values of $h_{\text{rss}}^{50\%}$ and $h_{\text{rss}}^{90\%}$ over the whole S6-VSR2/3 for the combined results (i.e.\ averaged over time) from all the networks. 
} 
\label{table:SGQ9}
\end{table}

\begin{table}
\centering
\begin{tabular}{cc|cc|cc}\hline\hline
   &      & \multicolumn{4}{c}{all networks} \\
 $f$&  $\tau$  &\multicolumn{2}{c|}{Linear}  & \multicolumn{2}{c}{Circular} \\ 
$[$Hz$]$ & $[$ms$]$  & $h_{\text{rss}}^{50\%}$ & $h_{\text{rss}}^{90\%}$ & $h_{\text{rss}}^{50\%}$ & $h_{\text{rss}}^{90\%}$ \\ \hline
2000    &   1.0   &  47.3 & 288 & 34.8 & 78.9 \\
2090    &   200   &  42.9 & 218 & 31.7 & 66.0 \\
2590    &   200   &  52.2 & 255 & 39.1 & 79.5 \\
3067    &   0.65  &  91.9 & 546 & 72.9 & 569 \\
\hline\hline
\end{tabular}
\caption{Values of $h_{\text{rss}}^{50\%}$ and $h_{\text{rss}}^{90\%}$ 
(for 50\% and 90\% detection efficiency at the chosen thresholds of 1/(8 yr) per frequency band),
in units of $10^{-22} \, {\rm Hz}^{-1/2}$,
for linearly and circularly polarized ring-downs characterized by parameters $f$ and $\tau$.
 }
\label{table:RD}
\end{table}

\begin{table}
\centering
\begin{tabular}{ccl|c|cc}\hline\hline
$f_{\text{low}}$& $\Delta{f}$ & $\tau$  & H1L1V1   & \multicolumn{2}{c}{all networks } \\ 
$[$Hz$]$ & $[$Hz$]$ & $[$ms$]$  &   $h_{\text{rss}}^{50\%}$  &  $h_{\text{rss}}^{50\%}$ & $h_{\text{rss}}^{90\%}$ \\ \hline
100 & 100 & 100 & 8.1 & 11.5 & 91.2 \\
250 & 100 & 100 & 7.5 & 10.5 & 43.1 \\
1000 & 10 & 100 & 15.5 & 22.5 & 93.6 \\
1000 & 1000 & 10 & 30.5 & 39.7 & 130 \\
1000 & 1000 & 100 & 76.8 & 76.7 & 492 \\

2000 & 100 & 100 & 35.7 & 40.3 & 193\\
2000 & 1000 & 10 & 55.6 & 63.1 & 211\\
3500 & 100 & 100 & 71.8 & 90.3 & 332\\
3500 & 1000 & 10 & 114 & 125 & 371\\

\hline\hline
\end{tabular}
\caption{Values of $h_{\text{rss}}^{50\%}$ and $h_{\text{rss}}^{90\%}$ 
(for 50\% and 90\% detection efficiency at the chosen thresholds of 1/(8 yr) per frequency band), in units of 
$10^{-22}$ ${\rm Hz}^{-1/2}$, for band-limited white noise waveforms 
characterized by parameters $f_{\text{low}}$, $\Delta{f}$, and $\tau$.
}
\label{table:WNB}
\end{table}

\subsection{Systematic Uncertainties}

The most relevant systematic uncertainty in the astrophysical interpretation of our results is due to the calibration error on the strain data produced by each detector ~\cite{LIGOS5,VirgoS2}.
The effect of calibration systematics on network detection efficiency has been 
estimated by dedicated simulations of GW signals in which the signal amplitude and 
phase at each detector is randomly jittered according to the modeled distribution of 
calibration errors for that detector.
The resulting network detection efficiency marginalizes the effect of the systematic 
uncertainties over the observation time. The main effect can be parametrized as an overall
shift of the detection efficiency curves along the signal strength axis. The largest 
effect over the injected signal waveforms was a $8\%$ increase of the $\hrss$ amplitude at 
fixed detection efficiency \footnote{Note that, due to an incomplete knowledge of the actuation resonances in [3000,4000] Hz band of the Hanford detector, very conservative assumptions have been made on calibration uncertainties; the networks including H1 in that frequency band feature a large efficiency loss due to calibration systematics of $24\%$. }. 
To produce the astrophysical limits
shown in Section \ref{sec:results}, we use the reduced detection efficiency curves obtained by shifting the original 
fits from Subsection \ref{sec:simulations} and the results in Tables \ref{table:SGQ9}, \ref{table:RD}, and \ref{table:WNB} to $8\%$ larger $\hrss$ values.

\section{Search Results}\label{sec:results}

The on-source data have been analyzed following the procedures tuned 
through the investigation of the off-source background sample, as described in 
Appendix \ref{sec:MIFAR}. 
No on-source event has been found above the threshold false alarm rate of once in 8 years 
per frequency band, and the distribution of on-source events is in agreement with 
the measured background. Table \ref{tab:LoudestIFAR} lists the five most significant 
on-source events, as ranked by their Inverse False Alarm Rate (IFAR = 1/FAR), and 
taking into account the trial factor due to the three independent searches performed on 
the disjoint frequency bands.  
 
In addition to the events reported in Table \ref{tab:LoudestIFAR}, this 
search also detected an on-source event showing a chirping waveform 
compatible with a compact binary coalescence at a signal-to-noise ratio 
$\sim 17$ in the H1L1V1 network.  This event was first identified by a 
low-latency burst search within minutes of its occurrence on 
September 16, 2010 and was thoroughly investigated in follow-up studies. 
Its Inverse False Alarm Rate was estimated at 1.1 yr from 
comparison with the burst reference background over all trials. After 
the completion of the analysis, this event was revealed to be a blind 
hardware injection \cite{GW100916opendata} intended as an end-to-end test of the search for 
transient signals \footnote{ Signal injections were performed via direct excitation of 
the interferometer mirror test masses. Some of these {\it hardware injections} were 
intended to mimic a coherent GW excitation across the network 
and to provide an end-to-end verification of the 
detector instrumentation, the data acquisition system and the data analysis software.
In addition to those, a {\it blind injection challenge} was 
realized consisting of injecting a few simulated signals at times not announced to the 
collaborations. This was done for the purpose of testing the data analysis pipelines and event validation protocols.}.
As such, the event was removed from the final results.

\begin{table}[htbp]
\begin{tabular}{|c|c|c|c|c|}
\hline
IFAR [yr]& freq. band & network & SNR & FAP \\
\hline
0.64 & 0.2-1.6 kHz & H1L1  & 11 & 0.59\\
\hline
0.36 & 64-200 Hz & H1L1V1 & 19 & 0.47\\
\hline
0.28 & 0.2-1.6 kHz & H1L1 & 12  &  0.33   \\
\hline
0.19 & 0.2-1.6 kHz & H1L1 & 10 & 0.35\\
\hline
0.17 & 1.6-5 kHz & H1V1 & 9 & 0.24\\
\hline
\end{tabular}
\caption{The five most significant events present in the on-source data. 
IFAR is the Inverse False Alarm Rate [yr] of the event in the entire search, 
SNR is the signal-to-noise ratio in the whole network, and FAP is the false 
alarm probability (probability of getting at least as many accidental events 
as those observed with IFAR $\geq$ the value reported in the first column).}
\label{tab:LoudestIFAR}
\end{table}

\subsection{Upper Limits}
\label{sec:ULs}

The new null result can be combined with the previous ones from the latest scientific runs by LIGO and Virgo \cite{Abbott:2009zi, abbott-2009b, S5VSR1Burst}  to 
complete the results achieved by initial generation interferometric detectors.

Assuming a Poisson distribution of astrophysical sources and in the special case of no surviving candidate events,
the 90\% confidence upper limit is computed as

\begin{equation}\label{eqn:ul}
R_{90\%}=\frac{2.3}{\sum_k \epsilon_k T_k}\,,
\end{equation}
where 2.3 = - log(1 - 0.9), $\epsilon_k$ and $T_k$ are respectively the detection efficiency and the observation time of the different network configurations in homogeneous sub-periods of observation $k$ \cite{Sutton:2010}, including all available LIGO and LIGO--Virgo observations since November 2005~\cite{Abbott:2009zi, abbott-2009b, S5VSR1Burst}.  

\begin{figure}
\begin{center}
\includegraphics*[width=0.5\textwidth]{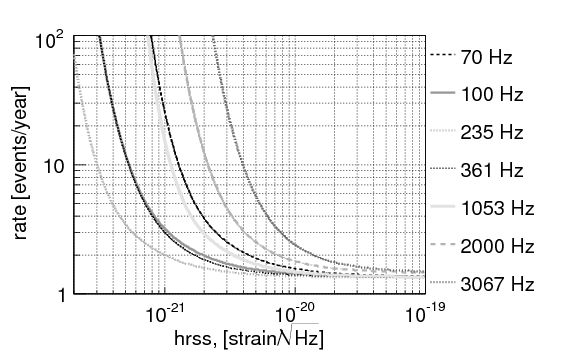}

\caption{Upper limits at 90\% confidence on the rate of gravitational-wave bursts at Earth as a function
of $\hrss$ signal amplitude for selected sine-Gaussian waveforms with $Q=9$. The results include all the 
LIGO and LIGO--Virgo observations since November 2005.}
\label{fig:S5-S6upperlimits}
\end{center}
\end{figure}

Figure~\ref{fig:S5-S6upperlimits} shows the upper limits on the rate of gravitational-wave bursts at the Earth as a function of signal strength ($\hrss$) for selected sine-Gaussian waveforms. 
The second joint LIGO--Virgo run increases the previous total observation time by roughly 50\%, totaling $1.74 \ \mathrm{yr}$. Therefore, the resulting 90\% upper limit on the rate for strong signals (asymptotic behaviour for $\epsilon_k\rightarrow 100\%$) decreases from $2.0$ to $1.3 \ \mathrm{yr^{-1}}$ for the 64 -- 1600 Hz band (from $2.2$ to $1.4 \ \mathrm{yr^{-1}}$ for the band above 1.6 kHz).

The results can also be interpreted as limits on the rate
density of GW bursts (number per year and per $\mathrm{Mpc^{3}}$) assuming a standard-candle
source, isotropically distributed, as previously reported in \cite{S5VSR1Burst}.  Denoting by
$h_0^2$ the average value of the GW squared amplitude $h_\mathrm{rss}^2$ 
at a fiducial distance $r_0$ from the source, the energy converted to GWs is 
\begin{equation}\label{eq:SGenergy}
E_{\text{GW}}= \frac{\pi^2 c^3}{G} \, r_0^2 \, f_0^2 \, h_0^2 \, .
\end{equation}
where $f_0$ is the central frequency of GW emission. 

Considering a population of standard-candle sources randomly 
oriented with respect to the Earth and at a distance $r_0$, we can interpret the 
$h_0^2$ as the average GW squared amplitude impinging on the Earth (e.g. averaged over source parameters 
such as inclination angle). Equation \ref{eq:SGenergy} can then be used to estimate $h_0(E_{\mathrm{GW}},f_0,r_0)$ and 
in particular sets the inverse proportionality between the average $h_\mathrm{rss}$ at Earth 
and source distance $r$: $h r = h_0 r_0$.
Assuming a uniform distribution in the sky and in time of these standard-candle sources, the expected rate of detections is
\begin{eqnarray}
N_\mathrm{det} & = & 4\pi {\cal R} T \int^{\infty}_{0}dr \, r^2 \epsilon(r) \nonumber \\
& = & 4\pi {\cal R} T (h_0 r_0)^3 \int^{\infty}_{0}dh  \, h^{-4} \epsilon(h) .
\end{eqnarray}
where ${\cal R}$ is the rate density of the standard-candle sources, 
$T$ the overall observation time, and $\epsilon(h)$ the detection efficiency as measured by our simulations.

Hence, the 90\%
confidence upper limit on rate density ${\cal R}$ of such standard-candle sources is
\begin{eqnarray}
{\cal R}_\mathrm{90\%} = \frac{2.3}{4\pi T (h_0 r_0)^3 \int_0^\infty  \!\! dh \, h^{-4}
\epsilon(h)} \, .
\end{eqnarray}
The resulting ${\cal R}_\mathrm{90\%}$ is dominated by the part of the detection efficiency curve at small GW amplitude $h$. 
Due to the relative orientations of the LIGO-Virgo detectors, detection efficiency curves for linearly polarized sine-Gaussian waveforms are approximately the same of those for elliptically polarized ones; the numerical values of ${\cal R}_\mathrm{90\%}$ are close within a few percent for both source models.

\begin{figure}
\begin{center}
\mbox{\includegraphics*[width=0.5\textwidth]{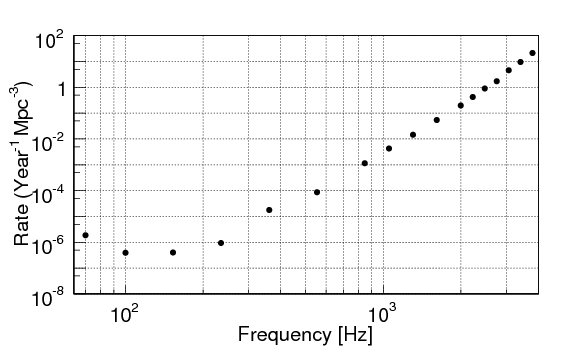}}
\caption{Rate limit per unit volume for standard-candle sources at the 90\% confidence level for a
linearly polarized sine-Gaussian standard-candle with $E_{\text{GW}}=M_\odot{c^2}$. Within an accuracy of a few percent, the same numerical results hold also for sources emitting circularly polarized GWs, 
which would subsequently appear elliptically polarized at the Earth.
In this Figure, all LIGO and LIGO--Virgo observations since November 2005 have been combined together. 
}
\label{fig:isotropicUL}
\end{center}
\end{figure}

Figure~\ref{fig:isotropicUL} shows the rate density upper limits of sources as a function of
frequency. This result can be interpreted in the following way: given a standard-candle source
with a characteristic frequency $f$ and energy $E_{\mathrm{GW}}$,
the corresponding rate limit is 
${\cal R}_\mathrm{90\%}(f) (M_\odot c^2 / E_\mathrm{GW})^{3/2}~\mathrm{yr}^{-1}\mathrm{Mpc}^{-3}$.

\begin{figure}
\begin{center}
\mbox{\includegraphics*[width=0.5\textwidth]{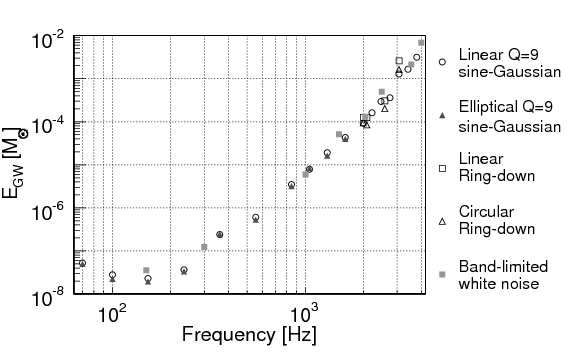}}
\caption{Typical GW energy in solar masses at 50\% detection efficiency for standard-candle sources emitting 
 at 10 kpc for the waveforms listed in Tables \ref{table:SGQ9}, \ref{table:RD}, and \ref{table:WNB} considering 
the H1L1V1 network and the LIGO-Virgo observations since July 2009. 
}
\label{fig:mass}
\end{center}
\end{figure}

The typical GW energy in units of solar masses for LIGO-Virgo
observation is shown in Figure \ref{fig:mass}
computed with Equation \ref{eq:SGenergy} using the measured $h_{rss}$ at 50\% detection efficiency for the 
tested waveforms assuming a standard candle source emitting 
at a distance of 10 kpc.
The mass scales with the square of the fiducial distance and the results are robust over the very wide class 
of waveforms tested. 
As expected, the GW energy is strongly dependent on the spectral sensitivity of the network, with a negligible dependence on the specific waveform characteristics. 

\section{Summary and Discussion}\label{sec:summaries}

This paper reports the results achieved by the LIGO and Virgo detectors in the search
for GW transients of duration $\lesssim 1$ s, without
assumptions on the signal waveform, polarization, direction or arrival time. 

Three detectors were operating at the Hanford, Livingston and Pisa sites 
during the second joint observation of LIGO and Virgo in 2009-2010. 
The detectors implemented hardware upgrades in order to prototype new 
subsystems planned for the upcoming advanced detectors. 
The resulting sensitivities to GWs were comparable to those achieved 
during the first LIGO-Virgo run. The main contribution 
of the second run is a $50\%$ increase in accumulated observation time.

No event candidates were found in this search. 
We set better upper limits on the rate of gravitational-wave bursts at Earth 
and on the rate density of burst sources per unit time and volume. These limits combine
all available information from the LIGO--Virgo joint runs and set the state-of-the-art on 
all-sky searches for transient gravitational waves of short duration.

The reported $h_\text{rss}$ amplitude of the GW at Earth can be converted into the 
energy emitted by a source at some fiducial distance assuming a simple model 
as in Equation \ref{eq:SGenergy}. For example, the 
energy emitted in gravitational waves in units of solar masses at a distance of 10 kpc 
and considering measured $h_{rss}$ at 50\% detection efficiency 
(Table \ref{table:SGQ9}) 
is $\simeq 2.2 \cdot 10^{-8} M_{\odot}$ 
for signal frequencies near 150 Hz  ($5.6 \cdot 10^{-2} M_{\odot}$ at 16 Mpc).
These GW energies, though obviously depending on the signal frequency, are 
approximately constant over different polarization models of the GW emission, 
including linearly polarized sources, circularly polarized sources and 
un-polarized emission with random polarization amplitudes (see Tables 
\ref{table:SGQ9}, \ref{table:RD}, and \ref{table:WNB}).

The long baseline interferometric detectors LIGO and Virgo are currently 
being upgraded to their advanced configurations, and the next joint observation 
is planned for 2015. Another advanced detector, LCGT \cite{uchiyama2004present, kuroda2010status}, is being built 
in Japan, and there are proposals to realize an additional advanced LIGO 
detector outside the USA.  These advanced detectors should achieve strain
sensitivities a factor of ten better than the first-generation detectors.
For example, at design sensitivity these detectors should detect a
typical core-collapse supernova anywhere in the galaxy~\cite{0264-9381-27-19-194005}
and will be able to put constraints on extreme scenarios for core collapse supernovae within the Local Group~\cite{ott2008,2011LRR}. 
Other possible short duration sources, such as the merger of very high mass stellar black hole binaries, could be visible at distances exceeding 1 Gpc. 
During advanced detector observations, gravitational-wave detections are 
predicted to occur on a regular basis ~\cite{ratesdoc}, 
thus greatly expanding the field of gravitational-wave astrophysics.

%
%


\begin{acknowledgments}\label{sec:acknowledgments}

The authors gratefully acknowledge the support of the United States National 
Science Foundation for the construction and operation of the LIGO Laboratory, 
the Science and Technology Facilities Council of the United Kingdom, the 
Max-Planck-Society and the State of Niedersachsen/Germany for support of the 
construction and operation of the GEO\,600 detector, 
and the Italian Istituto Nazionale di Fisica Nucleare and the French Centre 
National de la Recherche Scientifique for the construction and operation of 
the Virgo detector. The authors also gratefully acknowledge the support of 
the research by these agencies and by the Australian Research Council, the 
Council of Scientific and Industrial Research of India, the Istituto 
Nazionale di Fisica Nucleare of Italy, the Spanish Ministerio de Educaci\'on 
y Ciencia, the Conselleria d'Economia Hisenda i Innovaci\'o of the Govern de 
les Illes Balears, the Foundation for Fundamental Research on Matter supported 
by the Netherlands Organisation for Scientific Research,
the Polish Ministry of Science and Higher Education, the FOCUS Programme
of Foundation for Polish Science, the Royal Society, 
the Scottish Funding Council, the Scottish Universities Physics Alliance, 
the National Aeronautics and Space Administration, the Carnegie Trust, 
the Leverhulme Trust, the David and Lucile Packard Foundation, the Research 
Corporation, and the Alfred P. Sloan Foundation. 
This document has been assigned LIGO Laboratory document number \ligodoc.

\end{acknowledgments}

\appendix

\section{Data Quality Flags}\label{sec:dataquality}

Data Quality Flags (DQFs) are intended to indicate periods of data taking which suffer from environmental and instrumental effects inducing noise into the data \cite{VirgoDetChar}. We followed the DQF strategy used in previous searches \cite{Abbott:2009zi, abbott-2009b,S5VSR1Burst}, organizing DQFs into 3 categories.
The different categories  reflect the level of understanding of the
detectors' performances as well as of the relation between
disturbances in the data set and environmental or instrumental causes. 

Category 1 DQFs mark segments of time (typically more than tens of seconds) 
when disturbances make analysis unfeasible. 
Data segments remaining after their application are used in the analysis.

Category 2 DQFs are connected to well-understood short duration (typically a few seconds) periods of noise transients. Data segments flagged by this 
category can be used for data conditioning and noise property estimation, but events emerging from these periods are discarded as very likely 
originating from the detector environment.

Finally, Category 3 DQFs denote periods that are only weakly correlated to environmental and instrumental monitors.
Such cuts are not reliable enough to be used as unconditional cuts. When applied to events generated by the search algorithm, they would reject a significant fraction (in extreme cases up to 15-20\%) of data. 
Their use is limited to significance calculations using the MIFAR statistic (see Appendix~\ref{sec:MIFAR}).

\section{Event-by-event vetoes}\label{sec:vetoes}

Often, GW candidate events identified in the on-source time can be linked to disturbances propagating through the detector from the environment or within the detector itself.
Our procedure for identifying such {\it event-by-event } vetoes in S6 and VSR2/3
follows that used in S5 and VSR1~\cite{Abbott:2009zi,S5VSR1Burst}. 
The GW channel and a large number of auxiliary channels are processed with the 
Kleine-Welle \cite{2004CQG21S1809C} algorithm, which looks for excess power transients. 
A hierarchical method \cite{hveto} is used to rank the statistical relationship between the transients found in the auxiliary channels and those found in the detector output. Based on these rankings, vetoes are defined for suspected noise events. Another veto used was based on significant statistical association of events observed in the GW channel and the auxiliary channels \cite{UPV} .

An additional set of Category 3 vetoes \cite{ballinger2009powerful} are applied to
events emerging from networks including Virgo; vetoes from this set are based on 
detector read-out channels which are known to be insensitive to gravitational waves.

Procedurally, the event-by-event vetoes are applied with the same conditions 
as their corresponding Category of data quality flags described in 
Appendix~\ref{sec:dataquality}.
  
\section{Detector networks and live times}\label{sec:Ltimes}

The total observation time for the analysis has been divided into four sub-periods (labeled A, B, C and D), separated by planned commissioning 
and upgrade breaks which changed the performance of the detectors. 
Table~\ref{tab:data}  
shows the observation time of each network configuration after the application 
of Category 1 and 2 DQFs. These times are not overlapping.
During the period from January to June 2010 (sub-period C), Virgo did not participate in the run 
because of hardware upgrades.

\begin{table}[htbp]
\begin{tabular}{|c|c|c|c|c|c|}
\hline
  detectors    &   A [days]   & B [days]   & C [days]& D [days] & TOT [days]\\
\hline
 H1L1V1      &   10.6     & 16.7    & -      & 24.9    &  52.2 \\
\hline\
 H1L1        &   -        &  6.2    & 51.4   & 26.8    &  84.5\\
 L1V1        &   10.2     & 10.7    & -      &  8.1    &  28.9 \\
 H1V1        &   12.6     & 21.3    & -      &  7.1    &  41.0\\
 \hline
  TOT        &   33.4     & 54.8    & 51.4   & 66.9    & 206.6\\

\hline
\end{tabular}
\caption{\label{tab:data}  Observation time for each detector
configuration after application of Category 1 and 2 DQFs for 
the four sub-periods A, B, C, and D. For period A, the observation time 
of the H1L1 network after subtracting the H1L1V1 observation time is 
negligible ($\sim 1$ day).
During period C, Virgo did not participate in the run.}
\end{table}

\section{Modified Inverse False Alarm Rate (MIFAR)}\label{sec:MIFAR}

We introduce the Modified Inverse False Alarm Rate (MIFAR) to account for the 
effect of Category 3 DQFs on the background.

Category 3 DQFs indicate a weak statistical correlation of the GW data with 
environmental and instrumental noise sources, and thus were used only as a 
cautionary tag when examining an event in candidate follow-ups. Moreover, 
the effectiveness of these flags is not constant between different sub-periods, 
network configurations or frequency bands. The use of Category 3 data quality 
as a tag allows us to produce two sets of events: the ``raw'' set (polluted to
some extent by noise glitches) and the subset of those events that are not tagged, the 
``clean'' set (with reduced observation time). 

In order to account for the difference in background distributions when 
assessing the significance of candidate GW events from the raw and clean 
sets, we use the following procedure: 
\begin{enumerate}
\item 
Within each homogeneous analysis (same detector's configuration, same 
tuning of analysis, same frequency band), 
we rank events from the two sets separately by their coherent network 
amplitude $\eta$; {\it i.e.}, if the event candidate is flagged by 
Category 3 data quality, it is ranked against the raw set of events, 
otherwise it is ranked against the clean set.
\item
Each event is then assigned a MIFAR as 
the inverse of the rate of higher-ranked background events in that set, {\it i.e.} the MIFAR is the 
IFAR of the event considering only that set. 
\item
We merge the events from the raw and clean sets into a single list, 
sorted by the MIFAR. For events with 
equal MIFAR the one with larger $\eta$ is ranked higher. This ranking is performed 
separately for each homogeneous analysis. 
\item
According to this merged ranking, we measure the IFAR of the events as the rate of the corresponding background event  with equal MIFAR. This measured IFAR is used as our ``universal'' ranking for all events in all analyses.
\item The final IFAR of any event over the entire search is just 1/3 of the value estimated in 
the previous step because of the trials factor: three independent analyses have been performed 
for the three disjoint frequency bands. No contribution to the trials factor comes from the analyses of
different detectors' configuration since these were performed on non-overlapped observation times.
\end{enumerate}

In each homogeneous analysis, setting a threshold on IFAR corresponds to two thresholds on $\eta$, one for the raw set and one for the clean data set.
Table~\ref{Tab:tuning} reports the  selected thresholds. These thresholds were tuned using the background and injection events, 
without considering the on-source events to avoid bias in candidate 
selection.

\begin{table*}[htb]
\begin{tabular}{|cc|ccc|ccc|ccc|}
\hline
& & \multicolumn{9}{c|}{Frequency Band [Hz]}\\
\multicolumn{2}{|c|}{network} & \multicolumn{3}{c|}{64--200} & \multicolumn{3}{c|}{200--2000} & \multicolumn{3}{c|}{2000--5000} \\
& & $cc$ & $\eta_1$ & $\eta_2$ & $cc$ & $\eta_1$ & $\eta_2$ & $cc$ & $\eta_1$ & $\eta_2$\\
\hline
       & A & 0.70 & 5.9 &   - & 0.70 & 6.0 &   - & 0.65 & 4.6 &   -\\
H1L1V1 & B & 0.60 & 6.5 & 6.5 & 0.60 & 4.6 & 3.9 & 0.60 & 3.9 &   -\\
       & D & 0.60 & 6.7 &4.7  & 0.60 & 4.5 & 4.3 & 0.60 & 4.7 & 4.4\\
\hline
\multirow{5}{*}{H1L1} & A & 0.65 &32.0 &   - & 0.65 & 7.4 &   - & 0.65 & 5.8 &   -\\
       & B & 0.60 & 8.5 & 8.1 & 0.60 & 5.5 & 5.5 & 0.60 & 4.7 &   -\\
       & C & 0.60 & 8.8 & 6.4 & 0.60 & 6.4 & 5.5 & 0.60 & 4.6 &   -\\
       & \multirow{2}{*}{D} & \multirow{2}{*}{0.60} & \multirow{2}{*}{8.9} & \multirow{2}{*}{8.9} & 0.60 &12.9 & 9.9 & \multirow{2}{*}{0.60} & \multirow{2}{*}{4.7} & \multirow{2}{*}{4.6}\\
       &   &      &     &     & 0.60 & 5.1 & 4.8 &      &     &    \\
\hline
       & A & 0.65 &16.8 &   - & 0.65 & 6.0 &   - & 0.65 & 5.5 &   -\\
L1V1   & B & 0.60 & 5.8 & 5.8 & 0.60 & 6.4 & 5.0 & 0.60 & 4.6 &   -\\
       & D & 0.60 &17.0 &17.0 & 0.60 &10.3 &10.3 & 0.60 & 7.3 & 6.9\\
\hline
       & A & 0.65 &10.2 &   - & 0.65 & 5.2 &   - & 0.65 & 5.4 &   -\\
H1V1   & B & 0.60 & 6.3 & 5.3 & 0.60 & 5.2 & 5.0 & 0.60 & 4.4 &   -\\
       & D & 0.60 & 9.1 & 9.0 & 0.60 & 6.1 & 6.1 & 0.60 & 6.3 & 6.2\\
\hline
\end{tabular}
\caption{\label{Tab:tuning} Thresholds per network, sub-period and frequency band for the homogeneous analyses performed. 
The threshold $\eta_{1}$ is applied to the events in the raw set (those 
in coincidence with a Category 3 DQF) and $\eta_{2}$ is applied to events 
in the clean set (not in coincidence with a DQF). These thresholds have been selected in order to ensure a IFAR $\geq$ 8 yr. }
\end{table*}


\begin{thebibliography}{47}
\expandafter\ifx\csname natexlab\endcsname\relax\def\natexlab#1{#1}\fi
\expandafter\ifx\csname bibnamefont\endcsname\relax
  \def\bibnamefont#1{#1}\fi
\expandafter\ifx\csname bibfnamefont\endcsname\relax
  \def\bibfnamefont#1{#1}\fi
\expandafter\ifx\csname citenamefont\endcsname\relax
  \def\citenamefont#1{#1}\fi
\expandafter\ifx\csname url\endcsname\relax
  \def\url#1{\texttt{#1}}\fi
\expandafter\ifx\csname urlprefix\endcsname\relax\def\urlprefix{URL }\fi
\providecommand{\bibinfo}[2]{#2}
\providecommand{\eprint}[2][]{\url{#2}}

\bibitem[{\citenamefont{Cutler and Thorne}(2002)}]{Cutler:2001}
\bibinfo{author}{\bibfnamefont{C.}~\bibnamefont{Cutler}} \bibnamefont{and}
  \bibinfo{author}{\bibfnamefont{K.~S.} \bibnamefont{Thorne}}, in
  \emph{\bibinfo{booktitle}{Proceedings of GR16}}, edited by
  \bibinfo{editor}{\bibfnamefont{N.~T.} \bibnamefont{Bishop}} \bibnamefont{and}
  \bibinfo{editor}{\bibfnamefont{S.~D.} \bibnamefont{Maharaj}}
  (\bibinfo{publisher}{WorldScientific}, \bibinfo{address}{Singapore},
  \bibinfo{year}{2002}), \eprint{gr-qc/0204090}.

\bibitem[{\citenamefont{{Pretorius}}(2009)}]{Pretorius:2007nq}
\bibinfo{author}{\bibfnamefont{F.}~\bibnamefont{{Pretorius}}}, in
  \emph{\bibinfo{booktitle}{Physics of Relativistic Objects in Compact
  Binaries: from Birth to Coalescence}}, edited by
  \bibinfo{editor}{\bibfnamefont{M.}~\bibnamefont{{Colpi}}}
  \bibnamefont{et~al.} (\bibinfo{publisher}{Springer Verlag, Canopus Publishing
  Limited}, \bibinfo{year}{2009}), \eprint{arXiv:0710.1338}.

\bibitem[{\citenamefont{Etienne et~al.}(2008)}]{Etienne:2007}
\bibinfo{author}{\bibfnamefont{Z.~B.} \bibnamefont{Etienne}}
  \bibnamefont{et~al.}, \bibinfo{journal}{Phys.~Rev.~D}
  \textbf{\bibinfo{volume}{77}}, \bibinfo{pages}{084002}
  (\bibinfo{year}{2008}), \eprint{arXiv:0712.2460}.

\bibitem[{\citenamefont{{Ott}}(2009)}]{ott2008}
\bibinfo{author}{\bibfnamefont{C.~D.} \bibnamefont{{Ott}}},
  \bibinfo{journal}{Class. Quantum Grav.} \textbf{\bibinfo{volume}{26}},
  \bibinfo{pages}{063001} (\bibinfo{year}{2009}).

\bibitem[{\citenamefont{Baiotti et~al.}(2007)}]{SNwave}
\bibinfo{author}{\bibfnamefont{L.}~\bibnamefont{Baiotti}} \bibnamefont{et~al.},
  \bibinfo{journal}{Class. Quantum Grav.} \textbf{\bibinfo{volume}{24}},
  \bibinfo{pages}{S187} (\bibinfo{year}{2007}).

\bibitem[{\citenamefont{{Mereghetti}}(2008)}]{SGR}
\bibinfo{author}{\bibfnamefont{S.}~\bibnamefont{{Mereghetti}}},
  \bibinfo{journal}{Astron. Astrophys. Rev.} \textbf{\bibinfo{volume}{15}},
  \bibinfo{pages}{225} (\bibinfo{year}{2008}).

\bibitem[{\citenamefont{Andersson and Comer}(2001)}]{glitches}
\bibinfo{author}{\bibfnamefont{N.}~\bibnamefont{Andersson}} \bibnamefont{and}
  \bibinfo{author}{\bibfnamefont{G.~L.} \bibnamefont{Comer}},
  \bibinfo{journal}{Phys. Rev. Lett.} \textbf{\bibinfo{volume}{87}},
  \bibinfo{pages}{241101} (\bibinfo{year}{2001}).

\bibitem[{\citenamefont{Damour and Vilenkin}(2001)}]{cusp}
\bibinfo{author}{\bibfnamefont{T.}~\bibnamefont{Damour}} \bibnamefont{and}
  \bibinfo{author}{\bibfnamefont{A.}~\bibnamefont{Vilenkin}},
  \bibinfo{journal}{Phys. Rev. D} \textbf{\bibinfo{volume}{64}},
  \bibinfo{pages}{064008} (\bibinfo{year}{2001}).

\bibitem[{\citenamefont{Abbott et~al.}(2009{\natexlab{a}})}]{Abbott:2007kv}
\bibinfo{author}{\bibfnamefont{B.}~\bibnamefont{Abbott}} \bibnamefont{et~al.}
  (\bibinfo{collaboration}{LIGO Scientific Collaboration}),
  \bibinfo{journal}{Rept.~Prog.~Phys.} \textbf{\bibinfo{volume}{72}},
  \bibinfo{pages}{076901} (\bibinfo{year}{2009}{\natexlab{a}}),
  \eprint{arXiv:0711.3041}.

\bibitem[{\citenamefont{Acernese et~al.}(2006)}]{Acernese2006}
\bibinfo{author}{\bibfnamefont{F.}~\bibnamefont{Acernese}}
  \bibnamefont{et~al.}, \bibinfo{journal}{Journal of Physics: Conference
  Series} \textbf{\bibinfo{volume}{32}}, \bibinfo{pages}{223}
  (\bibinfo{year}{2006}),
  \urlprefix\url{http://stacks.iop.org/1742-6596/32/i=1/a=033}.

\bibitem[{\citenamefont{Allen et~al.}(2000)}]{IGEC:2000}
\bibinfo{author}{\bibfnamefont{Z.~A.} \bibnamefont{Allen}} \bibnamefont{et~al.}
  (\bibinfo{collaboration}{International Gravitational Event Collaboration}),
  \bibinfo{journal}{Phys. Rev. Lett.} \textbf{\bibinfo{volume}{85}},
  \bibinfo{pages}{5046} (\bibinfo{year}{2000}).

\bibitem[{\citenamefont{Astone et~al.}(2003)}]{IGEC:2003}
\bibinfo{author}{\bibfnamefont{P.}~\bibnamefont{Astone}} \bibnamefont{et~al.}
  (\bibinfo{collaboration}{International Gravitational Event Collaboration}),
  \bibinfo{journal}{Phys.~Rev.~D} \textbf{\bibinfo{volume}{68}},
  \bibinfo{pages}{022001} (\bibinfo{year}{2003}), \eprint{astro-ph/0302482}.

\bibitem[{\citenamefont{Astone et~al.}(2007)\citenamefont{Astone, Babusci,
  Baggio, Bassan, Bignotto, Bonaldi, Camarda, Carelli, Cavallari, Cerdonio
  et~al.}}]{IGEC:2007}
\bibinfo{author}{\bibfnamefont{P.}~\bibnamefont{Astone}},
  \bibinfo{author}{\bibfnamefont{D.}~\bibnamefont{Babusci}},
  \bibinfo{author}{\bibfnamefont{L.}~\bibnamefont{Baggio}},
  \bibinfo{author}{\bibfnamefont{M.}~\bibnamefont{Bassan}},
  \bibinfo{author}{\bibfnamefont{M.}~\bibnamefont{Bignotto}},
  \bibinfo{author}{\bibfnamefont{M.}~\bibnamefont{Bonaldi}},
  \bibinfo{author}{\bibfnamefont{M.}~\bibnamefont{Camarda}},
  \bibinfo{author}{\bibfnamefont{P.}~\bibnamefont{Carelli}},
  \bibinfo{author}{\bibfnamefont{G.}~\bibnamefont{Cavallari}},
  \bibinfo{author}{\bibfnamefont{M.}~\bibnamefont{Cerdonio}},
  \bibnamefont{et~al.} (\bibinfo{collaboration}{IGEC-2 Collaboration}),
  \bibinfo{journal}{Phys. Rev. D} \textbf{\bibinfo{volume}{76}},
  \bibinfo{pages}{102001} (\bibinfo{year}{2007}),
  \urlprefix\url{http://link.aps.org/doi/10.1103/PhysRevD.76.102001}.

\bibitem[{\citenamefont{Astone et~al.}(2010)\citenamefont{Astone, Baggio,
  Bassan, Bignotto, Bonaldi, Bonifazi, Cavallari, Cerdonio, Coccia, Conti
  et~al.}}]{IGEC:2010}
\bibinfo{author}{\bibfnamefont{P.}~\bibnamefont{Astone}},
  \bibinfo{author}{\bibfnamefont{L.}~\bibnamefont{Baggio}},
  \bibinfo{author}{\bibfnamefont{M.}~\bibnamefont{Bassan}},
  \bibinfo{author}{\bibfnamefont{M.}~\bibnamefont{Bignotto}},
  \bibinfo{author}{\bibfnamefont{M.}~\bibnamefont{Bonaldi}},
  \bibinfo{author}{\bibfnamefont{P.}~\bibnamefont{Bonifazi}},
  \bibinfo{author}{\bibfnamefont{G.}~\bibnamefont{Cavallari}},
  \bibinfo{author}{\bibfnamefont{M.}~\bibnamefont{Cerdonio}},
  \bibinfo{author}{\bibfnamefont{E.}~\bibnamefont{Coccia}},
  \bibinfo{author}{\bibfnamefont{L.}~\bibnamefont{Conti}}, \bibnamefont{et~al.}
  (\bibinfo{collaboration}{IGEC-2 Collaboration}), \bibinfo{journal}{Phys. Rev.
  D} \textbf{\bibinfo{volume}{82}}, \bibinfo{pages}{022003}
  (\bibinfo{year}{2010}),
  \urlprefix\url{http://link.aps.org/doi/10.1103/PhysRevD.82.022003}.

\bibitem[{\citenamefont{Abbott et~al.}(2007)}]{LIGOS4burst}
\bibinfo{author}{\bibfnamefont{B.}~\bibnamefont{Abbott}} \bibnamefont{et~al.}
  (\bibinfo{collaboration}{{LIGO} Scientific Collaboration}),
  \bibinfo{journal}{Class. Quantum Grav.} \textbf{\bibinfo{volume}{24}},
  \bibinfo{pages}{5343} (\bibinfo{year}{2007}), \eprint{arXiv:0704.0943}.

\bibitem[{\citenamefont{Abbott et~al.}(2009{\natexlab{b}})}]{Abbott:2009zi}
\bibinfo{author}{\bibfnamefont{B.~P.} \bibnamefont{Abbott}}
  \bibnamefont{et~al.} (\bibinfo{collaboration}{LIGO Scientific
  Collaboration}), \bibinfo{journal}{\prd} \textbf{\bibinfo{volume}{80}},
  \bibinfo{pages}{102001} (\bibinfo{year}{2009}{\natexlab{b}}).

\bibitem[{\citenamefont{Abbott et~al.}(2009{\natexlab{c}})}]{abbott-2009b}
\bibinfo{author}{\bibfnamefont{B.~P.} \bibnamefont{Abbott}}
  \bibnamefont{et~al.} (\bibinfo{collaboration}{LIGO Scientific
  Collaboration}), \bibinfo{journal}{\prd} \textbf{\bibinfo{volume}{80}},
  \bibinfo{pages}{102002} (\bibinfo{year}{2009}{\natexlab{c}}).

\bibitem[{\citenamefont{Abadie et~al.}(2010{\natexlab{a}})}]{S5VSR1Burst}
\bibinfo{author}{\bibfnamefont{J.}~\bibnamefont{Abadie}} \bibnamefont{et~al.}
  (\bibinfo{collaboration}{LIGO Scientific Collaboration and Virgo
  Collaboration}), \bibinfo{journal}{Phys. Rev. D}
  \textbf{\bibinfo{volume}{81}}, \bibinfo{pages}{102001}
  (\bibinfo{year}{2010}{\natexlab{a}}).

\bibitem[{\citenamefont{{The LIGO Scientific Collaboration}
  et~al.}(2011)\citenamefont{{The LIGO Scientific Collaboration}, {Virgo
  Collaboration: J.~Abadie}, {Abbott}, {Abbott}, {Abbott}, {Abernathy},
  {Accadia}, {Acernese}, {Adams}, {Adhikari} et~al.}}]{2011arXiv1109.3498T}
\bibinfo{author}{\bibnamefont{{The LIGO Scientific Collaboration}}},
  \bibinfo{author}{\bibnamefont{{Virgo Collaboration: J.~Abadie}}},
  \bibinfo{author}{\bibfnamefont{B.~P.} \bibnamefont{{Abbott}}},
  \bibinfo{author}{\bibfnamefont{R.}~\bibnamefont{{Abbott}}},
  \bibinfo{author}{\bibfnamefont{T.~D.} \bibnamefont{{Abbott}}},
  \bibinfo{author}{\bibfnamefont{M.}~\bibnamefont{{Abernathy}}},
  \bibinfo{author}{\bibfnamefont{T.}~\bibnamefont{{Accadia}}},
  \bibinfo{author}{\bibfnamefont{F.}~\bibnamefont{{Acernese}}},
  \bibinfo{author}{\bibfnamefont{C.}~\bibnamefont{{Adams}}},
  \bibinfo{author}{\bibfnamefont{R.}~\bibnamefont{{Adhikari}}},
  \bibnamefont{et~al.}, \bibinfo{journal}{ArXiv e-prints}
  (\bibinfo{year}{2011}), \eprint{1109.3498}.

\bibitem[{\citenamefont{{Gehrels} et~al.}(2004)\citenamefont{{Gehrels},
  {Chincarini}, {Giommi}, {Mason}, {Nousek}, {Wells}, {White}, {Barthelmy},
  {Burrows}, {Cominsky} et~al.}}]{swift04}
\bibinfo{author}{\bibfnamefont{N.}~\bibnamefont{{Gehrels}}},
  \bibinfo{author}{\bibfnamefont{G.}~\bibnamefont{{Chincarini}}},
  \bibinfo{author}{\bibfnamefont{P.}~\bibnamefont{{Giommi}}},
  \bibinfo{author}{\bibfnamefont{K.~O.} \bibnamefont{{Mason}}},
  \bibinfo{author}{\bibfnamefont{J.~A.} \bibnamefont{{Nousek}}},
  \bibinfo{author}{\bibfnamefont{A.~A.} \bibnamefont{{Wells}}},
  \bibinfo{author}{\bibfnamefont{N.~E.} \bibnamefont{{White}}},
  \bibinfo{author}{\bibfnamefont{S.~D.} \bibnamefont{{Barthelmy}}},
  \bibinfo{author}{\bibfnamefont{D.~N.} \bibnamefont{{Burrows}}},
  \bibinfo{author}{\bibfnamefont{L.~R.} \bibnamefont{{Cominsky}}},
  \bibnamefont{et~al.}, \bibinfo{journal}{\apj} \textbf{\bibinfo{volume}{611}},
  \bibinfo{pages}{1005} (\bibinfo{year}{2004}).

\bibitem[{\citenamefont{{Gehrels} et~al.}(2005)\citenamefont{{Gehrels},
  {Chincarini}, {Giommi}, {Mason}, {Nousek}, {Wells}, {White}, {Barthelmy},
  {Burrows}, {Cominsky} et~al.}}]{2005ApJ...621..558G}
\bibinfo{author}{\bibfnamefont{N.}~\bibnamefont{{Gehrels}}},
  \bibinfo{author}{\bibfnamefont{G.}~\bibnamefont{{Chincarini}}},
  \bibinfo{author}{\bibfnamefont{P.}~\bibnamefont{{Giommi}}},
  \bibinfo{author}{\bibfnamefont{K.~O.} \bibnamefont{{Mason}}},
  \bibinfo{author}{\bibfnamefont{J.~A.} \bibnamefont{{Nousek}}},
  \bibinfo{author}{\bibfnamefont{A.~A.} \bibnamefont{{Wells}}},
  \bibinfo{author}{\bibfnamefont{N.~E.} \bibnamefont{{White}}},
  \bibinfo{author}{\bibfnamefont{S.~D.} \bibnamefont{{Barthelmy}}},
  \bibinfo{author}{\bibfnamefont{D.~N.} \bibnamefont{{Burrows}}},
  \bibinfo{author}{\bibfnamefont{L.~R.} \bibnamefont{{Cominsky}}},
  \bibnamefont{et~al.}, \bibinfo{journal}{\apj} \textbf{\bibinfo{volume}{621}},
  \bibinfo{pages}{558} (\bibinfo{year}{2005}).

\bibitem[{\citenamefont{Abadie et~al.}(2010{\natexlab{b}})}]{S5LowMassLV}
\bibinfo{author}{\bibfnamefont{J.}~\bibnamefont{Abadie}} \bibnamefont{et~al.}
  (\bibinfo{collaboration}{LIGO Scientific Collaboration and Virgo
  Collaboration}), \bibinfo{journal}{\prd} \textbf{\bibinfo{volume}{82}},
  \bibinfo{pages}{102001} (\bibinfo{year}{2010}{\natexlab{b}}),
  \eprint{1005.4655}.

\bibitem[{\citenamefont{Klimenko et~al.}(2011)\citenamefont{Klimenko, Vedovato,
  Drago, Mazzolo, Mitselmakher, Pankow, Prodi, Re, Salemi, and
  Yakushin}}]{cWB-PRCmethod2011}
\bibinfo{author}{\bibfnamefont{S.}~\bibnamefont{Klimenko}},
  \bibinfo{author}{\bibfnamefont{G.}~\bibnamefont{Vedovato}},
  \bibinfo{author}{\bibfnamefont{M.}~\bibnamefont{Drago}},
  \bibinfo{author}{\bibfnamefont{G.}~\bibnamefont{Mazzolo}},
  \bibinfo{author}{\bibfnamefont{G.}~\bibnamefont{Mitselmakher}},
  \bibinfo{author}{\bibfnamefont{C.}~\bibnamefont{Pankow}},
  \bibinfo{author}{\bibfnamefont{G.}~\bibnamefont{Prodi}},
  \bibinfo{author}{\bibfnamefont{V.}~\bibnamefont{Re}},
  \bibinfo{author}{\bibfnamefont{F.}~\bibnamefont{Salemi}}, \bibnamefont{and}
  \bibinfo{author}{\bibfnamefont{I.}~\bibnamefont{Yakushin}},
  \bibinfo{journal}{Phys. Rev. D} \textbf{\bibinfo{volume}{83}},
  \bibinfo{pages}{102001} (\bibinfo{year}{2011}).

\bibitem[{\citenamefont{Schutz}(2011)}]{Schutz-networks:2011}
\bibinfo{author}{\bibfnamefont{B.~F.} \bibnamefont{Schutz}},
  \bibinfo{journal}{Classical and Quantum Gravity}
  \textbf{\bibinfo{volume}{28}}, \bibinfo{pages}{125023}
  (\bibinfo{year}{2011}),
  \urlprefix\url{http://stacks.iop.org/0264-9381/28/i=12/a=125023}.

\bibitem[{\citenamefont{Harry and the LIGO
  Scientific~Collaboration}(2010)}]{Harry-advLIGO:2010}
\bibinfo{author}{\bibfnamefont{G.~M.} \bibnamefont{Harry}} \bibnamefont{and}
  \bibinfo{author}{\bibnamefont{the LIGO Scientific~Collaboration}},
  \bibinfo{journal}{Classical and Quantum Gravity}
  \textbf{\bibinfo{volume}{27}}, \bibinfo{pages}{084006}
  (\bibinfo{year}{2010}),
  \urlprefix\url{http://stacks.iop.org/0264-9381/27/i=8/a=084006}.

\bibitem[{\citenamefont{Acernese et~al.}(2009)}]{Acernese-AdVreport:2009}
\bibinfo{author}{\bibfnamefont{F.}~\bibnamefont{Acernese}}
  \bibnamefont{et~al.}, \bibinfo{journal}{Report No. VIR0027A09}
  (\bibinfo{year}{2009}),
  \urlprefix\url{https://pub3.ego-gw.it/itf/tds/file.php?callFile=VIR-0027A-09%
.pdf}.

\bibitem[{\citenamefont{Adhikari et~al.}(2006)\citenamefont{Adhikari,
  Fritschel, and Waldman}}]{Adhikari:2006}
\bibinfo{author}{\bibfnamefont{R.}~\bibnamefont{Adhikari}},
  \bibinfo{author}{\bibfnamefont{P.}~\bibnamefont{Fritschel}},
  \bibnamefont{and} \bibinfo{author}{\bibfnamefont{S.}~\bibnamefont{Waldman}},
  \bibinfo{type}{Tech. Rep.} \bibinfo{number}{{LIGO}-T060156-v1},
  \bibinfo{institution}{{LIGO} Project} (\bibinfo{year}{2006}),
  \urlprefix\url{https://dcc.ligo.org/cgi-bin/DocDB/ShowDocument?docid=7384}.

\bibitem[{\citenamefont{Lorenzini}(2010)}]{lorenzini2010monolithic}
\bibinfo{author}{\bibfnamefont{M.}~\bibnamefont{Lorenzini}},
  \bibinfo{journal}{Classical and Quantum Gravity}
  \textbf{\bibinfo{volume}{27}}, \bibinfo{pages}{084021}
  (\bibinfo{year}{2010}).

\bibitem[{\citenamefont{Abadie et~al.}()}]{VirgoDetChar}
\bibinfo{author}{\bibfnamefont{J.}~\bibnamefont{Abadie}} \bibnamefont{et~al.}
  (\bibinfo{collaboration}{LIGO Scientific Collaboration and Virgo
  Collaboration}) (????).

\bibitem[{\citenamefont{Klimenko et~al.}(2008)}]{Klimenko:2008fu}
\bibinfo{author}{\bibfnamefont{S.}~\bibnamefont{Klimenko}}
  \bibnamefont{et~al.}, \bibinfo{journal}{Class. Quantum Grav.}
  \textbf{\bibinfo{volume}{25}}, \bibinfo{pages}{114029}
  (\bibinfo{year}{2008}), \eprint{arXiv:0802.3232}.

\bibitem[{\citenamefont{Abbott and et. al.}(2008)}]{Abbott:2008eh}
\bibinfo{author}{\bibfnamefont{B.}~\bibnamefont{Abbott}} \bibnamefont{and}
  \bibinfo{author}{\bibnamefont{et. al.}}, \bibinfo{journal}{Class. Quant.
  Grav.} \textbf{\bibinfo{volume}{25}}, \bibinfo{pages}{245008}
  (\bibinfo{year}{2008}).

\bibitem[{\citenamefont{Pankow et~al.}(2009)\citenamefont{Pankow, Klimenko,
  Mitselmakher, Yakushin, Vedovato, Drago, Mercer, and Ajith}}]{Pankow:2009lv}
\bibinfo{author}{\bibfnamefont{C.}~\bibnamefont{Pankow}},
  \bibinfo{author}{\bibfnamefont{S.}~\bibnamefont{Klimenko}},
  \bibinfo{author}{\bibfnamefont{G.}~\bibnamefont{Mitselmakher}},
  \bibinfo{author}{\bibfnamefont{I.}~\bibnamefont{Yakushin}},
  \bibinfo{author}{\bibfnamefont{G.}~\bibnamefont{Vedovato}},
  \bibinfo{author}{\bibfnamefont{M.}~\bibnamefont{Drago}},
  \bibinfo{author}{\bibfnamefont{R.~A.} \bibnamefont{Mercer}},
  \bibnamefont{and} \bibinfo{author}{\bibfnamefont{P.}~\bibnamefont{Ajith}},
  \bibinfo{journal}{Class. Quant. Grav.} \textbf{\bibinfo{volume}{26}},
  \bibinfo{pages}{204004} (\bibinfo{year}{2009}).

\bibitem[{\citenamefont{{Benhar} et~al.}(2004)\citenamefont{{Benhar},
  {Ferrari}, and {Gualtieri}}}]{Ferrari2004}
\bibinfo{author}{\bibfnamefont{O.}~\bibnamefont{{Benhar}}},
  \bibinfo{author}{\bibfnamefont{V.}~\bibnamefont{{Ferrari}}},
  \bibnamefont{and}
  \bibinfo{author}{\bibfnamefont{L.}~\bibnamefont{{Gualtieri}}},
  \bibinfo{journal}{\prd} \textbf{\bibinfo{volume}{70}},
  \bibinfo{pages}{124015} (\bibinfo{year}{2004}),
  \eprint{arXiv:astro-ph/0407529}.

\bibitem[{\citenamefont{Abadie et~al.}(2010{\natexlab{c}})}]{Abadie:2010xk}
\bibinfo{author}{\bibfnamefont{J.}~\bibnamefont{Abadie}} \bibnamefont{et~al.},
  \bibinfo{journal}{Phys. Rev. D} \textbf{\bibinfo{volume}{81}}
  (\bibinfo{year}{2010}{\natexlab{c}}).

\bibitem[{\citenamefont{Abadie et~al.}(2010{\natexlab{d}})}]{LIGOS5}
\bibinfo{author}{\bibfnamefont{J.}~\bibnamefont{Abadie}} \bibnamefont{et~al.}
  (\bibinfo{collaboration}{LIGO Scientific Collaboration}),
  \bibinfo{journal}{Nucl.~Instrum.~Meth.~A} \textbf{\bibinfo{volume}{624}},
  \bibinfo{pages}{223} (\bibinfo{year}{2010}{\natexlab{d}}),
  \eprint{arXiv:1007.3937}.

\bibitem[{\citenamefont{Accadia et~al.}(2011)}]{VirgoS2}
\bibinfo{author}{\bibfnamefont{T.}~\bibnamefont{Accadia}} \bibnamefont{et~al.}
  (\bibinfo{collaboration}{Virgo Collaboration}),
  \bibinfo{journal}{Class.~Quantum Grav.} \textbf{\bibinfo{volume}{28}},
  \bibinfo{pages}{025005} (\bibinfo{year}{2011}), \eprint{arXiv:1009.5190}.

\bibitem[{Open web page()}]{GW100916opendata}
Open web page, \urlprefix\url{http://www.ligo.org/science/GW100916/index.php}.

\bibitem[{\citenamefont{Sutton}(2010)}]{Sutton:2010}
\bibinfo{author}{\bibfnamefont{P.~J.} \bibnamefont{Sutton}},
  \bibinfo{journal}{Class. Quantum Grav.} \textbf{\bibinfo{volume}{26}},
  \bibinfo{pages}{245007} (\bibinfo{year}{2010}), \eprint{0905.4089v2}.

\bibitem[{\citenamefont{Uchiyama et~al.}(2004)\citenamefont{Uchiyama, Kuroda,
  Ohashi, Miyoki, Ishitsuka, Yamamoto, Hayakawa, Kasahara, Fujimoto, Kawamura
  et~al.}}]{uchiyama2004present}
\bibinfo{author}{\bibfnamefont{T.}~\bibnamefont{Uchiyama}},
  \bibinfo{author}{\bibfnamefont{K.}~\bibnamefont{Kuroda}},
  \bibinfo{author}{\bibfnamefont{M.}~\bibnamefont{Ohashi}},
  \bibinfo{author}{\bibfnamefont{S.}~\bibnamefont{Miyoki}},
  \bibinfo{author}{\bibfnamefont{H.}~\bibnamefont{Ishitsuka}},
  \bibinfo{author}{\bibfnamefont{K.}~\bibnamefont{Yamamoto}},
  \bibinfo{author}{\bibfnamefont{H.}~\bibnamefont{Hayakawa}},
  \bibinfo{author}{\bibfnamefont{K.}~\bibnamefont{Kasahara}},
  \bibinfo{author}{\bibfnamefont{M.}~\bibnamefont{Fujimoto}},
  \bibinfo{author}{\bibfnamefont{S.}~\bibnamefont{Kawamura}},
  \bibnamefont{et~al.}, \bibinfo{journal}{Classical and Quantum Gravity}
  \textbf{\bibinfo{volume}{21}}, \bibinfo{pages}{S1161} (\bibinfo{year}{2004}).

\bibitem[{\citenamefont{Kuroda}(2010)}]{kuroda2010status}
\bibinfo{author}{\bibfnamefont{K.}~\bibnamefont{Kuroda}},
  \bibinfo{journal}{Classical and Quantum Gravity}
  \textbf{\bibinfo{volume}{27}}, \bibinfo{pages}{084004}
  (\bibinfo{year}{2010}).

\bibitem[{\citenamefont{Yakunin et~al.}(2010)\citenamefont{Yakunin, Marronetti,
  Mezzacappa, Bruenn, Lee, Chertkow, Hix, Blondin, Lentz, Messer
  et~al.}}]{0264-9381-27-19-194005}
\bibinfo{author}{\bibfnamefont{K.~N.} \bibnamefont{Yakunin}},
  \bibinfo{author}{\bibfnamefont{P.}~\bibnamefont{Marronetti}},
  \bibinfo{author}{\bibfnamefont{A.}~\bibnamefont{Mezzacappa}},
  \bibinfo{author}{\bibfnamefont{S.~W.} \bibnamefont{Bruenn}},
  \bibinfo{author}{\bibfnamefont{C.-T.} \bibnamefont{Lee}},
  \bibinfo{author}{\bibfnamefont{M.~A.} \bibnamefont{Chertkow}},
  \bibinfo{author}{\bibfnamefont{W.~R.} \bibnamefont{Hix}},
  \bibinfo{author}{\bibfnamefont{J.~M.} \bibnamefont{Blondin}},
  \bibinfo{author}{\bibfnamefont{E.~J.} \bibnamefont{Lentz}},
  \bibinfo{author}{\bibfnamefont{O.~E.~B.} \bibnamefont{Messer}},
  \bibnamefont{et~al.}, \bibinfo{journal}{Classical and Quantum Gravity}
  \textbf{\bibinfo{volume}{27}}, \bibinfo{pages}{194005}
  (\bibinfo{year}{2010}).

\bibitem[{\citenamefont{{Fryer} and {New}}(2011)}]{2011LRR}
\bibinfo{author}{\bibfnamefont{C.~L.} \bibnamefont{{Fryer}}} \bibnamefont{and}
  \bibinfo{author}{\bibfnamefont{K.~C.~B.} \bibnamefont{{New}}},
  \bibinfo{journal}{Living Reviews in Relativity}
  \textbf{\bibinfo{volume}{14}}, \bibinfo{pages}{1} (\bibinfo{year}{2011}).

\bibitem[{\citenamefont{Abadie et~al.}(2010{\natexlab{e}})}]{ratesdoc}
\bibinfo{author}{\bibfnamefont{J.}~\bibnamefont{Abadie}} \bibnamefont{et~al.}
  (\bibinfo{collaboration}{LIGO Scientific Collaboration and Virgo
  Collaboration}), \bibinfo{journal}{Class. Quantum Grav.}
  \textbf{\bibinfo{volume}{27}}, \bibinfo{pages}{173001}
  (\bibinfo{year}{2010}{\natexlab{e}}).

\bibitem[{\citenamefont{{Chatterji} et~al.}(2004)\citenamefont{{Chatterji},
  {Blackburn}, {Martin}, and {Katsavounidis}}}]{2004CQG21S1809C}
\bibinfo{author}{\bibfnamefont{S.}~\bibnamefont{{Chatterji}}},
  \bibinfo{author}{\bibfnamefont{L.}~\bibnamefont{{Blackburn}}},
  \bibinfo{author}{\bibfnamefont{G.}~\bibnamefont{{Martin}}}, \bibnamefont{and}
  \bibinfo{author}{\bibfnamefont{E.}~\bibnamefont{{Katsavounidis}}},
  \bibinfo{journal}{Classical and Quantum Gravity}
  \textbf{\bibinfo{volume}{21}}, \bibinfo{pages}{1809} (\bibinfo{year}{2004}),
  \eprint{arXiv:gr-qc/0412119}.

\bibitem[{\citenamefont{{Smith} et~al.}(2011)\citenamefont{{Smith}, {Abbott},
  {Hirose}, {Leroy}, {MacLeod}, {McIver}, {Saulson}, and {Shawhan}}}]{hveto}
\bibinfo{author}{\bibfnamefont{J.~R.} \bibnamefont{{Smith}}},
  \bibinfo{author}{\bibfnamefont{T.}~\bibnamefont{{Abbott}}},
  \bibinfo{author}{\bibfnamefont{E.}~\bibnamefont{{Hirose}}},
  \bibinfo{author}{\bibfnamefont{N.}~\bibnamefont{{Leroy}}},
  \bibinfo{author}{\bibfnamefont{D.}~\bibnamefont{{MacLeod}}},
  \bibinfo{author}{\bibfnamefont{J.}~\bibnamefont{{McIver}}},
  \bibinfo{author}{\bibfnamefont{P.}~\bibnamefont{{Saulson}}},
  \bibnamefont{and}
  \bibinfo{author}{\bibfnamefont{P.}~\bibnamefont{{Shawhan}}},
  \bibinfo{journal}{Classical and Quantum Gravity}
  \textbf{\bibinfo{volume}{28}}, \bibinfo{pages}{235005}
  (\bibinfo{year}{2011}), \eprint{1107.2948}.

\bibitem[{\citenamefont{Isogai et~al.}(2010)}]{UPV}
\bibinfo{author}{\bibfnamefont{T.}~\bibnamefont{Isogai}} \bibnamefont{et~al.},
  in \emph{\bibinfo{booktitle}{Journal of Physics: Conference Series}}
  (\bibinfo{organization}{IOP Publishing}, \bibinfo{year}{2010}), vol.
  \bibinfo{volume}{243}, p. \bibinfo{pages}{012005}.

\bibitem[{\citenamefont{Ballinger}(2009)}]{ballinger2009powerful}
\bibinfo{author}{\bibfnamefont{T.}~\bibnamefont{Ballinger}},
  \bibinfo{journal}{Classical and Quantum Gravity}
  \textbf{\bibinfo{volume}{26}}, \bibinfo{pages}{204003}
  (\bibinfo{year}{2009}).

\end{thebibliography}

\end{document}